\documentclass[traditabstract]{aa}
\usepackage{graphicx}
\usepackage{txfonts}

\usepackage[sort]{natbib}
\bibpunct{(}{)}{;}{a}{}{,} 

\begin{document}

\title{Mass-radius relation of low and very low-mass stars revisited with the VLTI\thanks
{
       Based on data collected with the VLTI/VINCI and VLTI/AMBER instruments at ESO Paranal Observatory, programs ID 60.A-9220, 079.D-0565 and 080.D-0653.}}
 \subtitle{}

\author{      
	B.-O.   Demory\inst{1}
         \and D.  S\'egransan\inst{1}
         \and T.   Forveille\inst{2}
         \and D.   Queloz   \inst{1}
         \and J-L. Beuzit   \inst{2}
         \and X.   Delfosse \inst{2}
         \and E.   Di Folco \inst{1,3}
         \and P.   Kervella \inst{3}
         \and J.-B.~Le Bouquin \inst{4}
         \and C.   Perrier  \inst{2}
}

\offprints{Brice-Olivier Demory \email{Brice-Olivier.Demory@unige.ch}}

\institute{Observatoire de Gen\`eve, Universit\'e de Gen\`eve, 51 chemin des Maillettes, 1290 Sauverny, Suisse
      \and Laboratoire d'Astrophysique de Grenoble, 414 rue de la piscine, Domaine Universitaire de Saint Martin d'H\`eres, 38041 Grenoble, France
      \and LESIA, Observatoire de Paris, CNRS UMR 8109, UPMC, Universit\'e Paris Diderot, 5 place Jules Janssen, 92195 Meudon, France
      \and European Southern Observatory, Casilla 19001, Vitacura, Santiago 19, Chile
}

\date{Received  / Accepted }

\abstract{We measured the radii of 7 low and very low-mass stars using 
long baseline interferometry with the VLTI interferometer and its VINCI
and AMBER near-infrared recombiners. We use these new data, together 
with literature 
measurements, to examine the luminosity-radius and mass-radius relations 
for K and M dwarfs. The precision of the new interferometric radii 
now competes with what can be obtained for double-lined eclipsing 
binaries. Interferometry provides access to much less active stars,
as well as to stars with much better measured distances and luminosities,
and therefore complements the information obtained from eclipsing systems. 
The radii of magnetically quiet late-K to M dwarfs match the predictions
of stellar evolution models very well, providing direct confirmation 
that magnetic activity explains the discrepancy that was recently found 
for magnetically active eclipsing systems. 
The radii of the early K dwarfs 
are well reproduced for a mixing length parameter that approaches the
solar value, as qualitatively expected.

}

\maketitle

\section{Introduction}

Measuring accurate masses, radii, and luminosities, for low and very 
low-mass stars has always been observationally challenging. Precise 
individual masses and radii can be measured through combined 
photometric and spectroscopic observations of double-lined 
eclipsing binaries, but to date only $\sim$15 such systems are known 
with masses under 1~solar mass, and a few of those are too distant and 
faint for high precision work. Perhaps even more importantly, low 
mass eclipsing binaries tend, almost by construction, to be fast 
rotators and hence magnetically very active. Evolutionary models 
have been found to systematically underestimate the radii of 
very low mass eclipsing binaries \citep{Torres2002}, and the
uniformally high activity level of these objects is currently
the leading explanation for that discrepancy. \citet{Chabrier2007} 
find that increased surface spots coverage, and for partly convective
stars convection quenching by strong magnetic fields, can inflate 
the stellar radius by amounts which qualitatively
match the observed discrepancy.

Direct tests of that prediction, and validation of the 1-D structural 
models on objects which better match their assumptions, need 
radius measurements for slowly rotating and magnetically quiet 
very low-mass stars. Strong observational selection effects unfortunately 
ensure that all known eclipsing binaries have sufficiently short orbital 
periods that they are tidally synchronised, and therefore in turn that all 
are fast rotators. Measurements of magnetically quiet slow rotators
therefore have to use a different observing technique, long-baseline 
optical or infrared interferometry. \citet{Lane2001} and 
\citet{Segransan2003} both demonstrated 1-5\% radius precision
for stars that are only partially resolved.

The mass of these single, isolated, stars is not directly accessible,
and, strictly speaking, both \citet{Lane2001} and \citet{Segransan2003} 
therefore probe the luminosity-radius relation rather than the
mass-radius one. Accurate mass and luminosity measurements for
M dwarfs \citep[e.g.][]{Segransan2000} however demonstrate
that their K-band mass-luminosity relation has very low 
dispersion \citep{Delfosse2000}, and mass is therefore
largely interchangeable with absolute K-band luminosity.

Here we present direct angular diameter measurements for seven 
single K0.5 to M5.5 dwarfs, obtained with the VINCI and AMBER 
instruments on the Very Large Telescope Interferometer (VLTI) 
between 2003 and 2008. Section 2 describes those observations, 
the data analysis, and the angular diameter determination.
Section 3 discusses the luminosity-radius and mass-radius 
relation for very low-mass stars in the light of the new
measurements, and compares the empirical relations with 
theoretical predictions.

\section{Observations and data analysis}

\subsection{Sample}

The target list was largely determined by the capabilities of the VLTI
at the time of the observations. The limiting correlated magnitude of 
the low spectral resolution mode of AMBER on the 1.8m auxiliary telescopes
was K=4 in 2007, and improved to K=5.5 in 2008. We consequently selected
targets with apparent magnitudes between K=2.18 and K=4.38. Ongoing 
improvements to the VLTI infrastucture are expected to provide access
to fainter targets. We also computed the expected angular diameters
from flux-colour relations and literature photometry, and selected targets 
with a predicted diameter above 0.9~mas. This translates into a visibility 
of at most 0.8 on a 128~m baseline in the H-band, as needed to measure
diameters to a few \% uncertainty with the current amplitude calibration
precision of AMBER. One object, GJ\,879, was observed as a backup target 
during an unrelated observing program, and does not fulfill this minimum 
angular diameter specification.

\subsection{Observations}

\subsubsection{VINCI Observations}

GJ\,845\,A ($\epsilon$ Ind), GJ\,166\,A (DY Eri), GJ\,570\,A (KX Lib) 
and GJ\,663\,A were observed on the ESO Very Large Telescope 
Interferometer (VLTI) using its commissionning instrument, VINCI  
\citep{2000SPIE.4006...31K} with the two 35 cm test siderostats. 
Table \ref{table_obs} summarises the observation details. 
VINCI operated in the K-band and used single-mode optical fiber
couplers to recombine the light from two telescopes, and modulated 
the optical path difference around the white light fringe to 
produce interference fringes. This recombination scheme, first used 
in the FLUOR instrument \citep{1998SPIE.3350..856C}, produces high 
precision visibility values, thanks to the efficient conditioning
of the incoming wavefronts by the single mode fibers, to photometric 
monitoring of the light coupled into each input fibers, and to
fast scanning of the high quality fringes. \citep{KervellaSegransan2004}
extensively describe the data reduction for VINCI.\\

Our observing strategy alternated sequences of several hundred 
fringe scans on the target star and on spatially close calibrator 
stars, to efficiently sample the temporal and spatial structure 
of the atmospheric and instrumental transfer function. Adherence 
to this strategy could unfortunately not always be strict, since
scientific observations often had to give way to VLTI commissionning 
activities. For GJ\,166\,A, in particular, we could only keep three 
data points since all other measurements had no acceptably close
calibrator observations. That target additionally was observed
under poorer atmospheric conditions than all all other sources,
and with two calibrators respectively located 78 and 113 degrees 
away. It is by far our worst quality measurement, and is thus discarded from this study.

\subsubsection{AMBER Observations}

We used the AMBER (Astronomical Multi-BEam combineR) recombiner of the
VLTI to measure the radiii of GJ\,166\,A, GJ\,887, 
\textit{Proxima} (GJ\,551) and GJ\,879. Table \ref{table_obs}
summarizes the observing circumstances. AMBER uses single-mode 
fibers for wavefront filtering and produces spectrally-dispersed
fringes in the J, H, and K near-infrared bands \citep{Petrov2007}. \\
We used the 1.8m-diameter VLT auxilliary telescopes (AT) on the A0-K0-G1 
baseline triplet, which at the time of our observations offered the 
longest available baselines and therefore the highest angular resolution. 
We selected the low spectral resolution mode (low-JHK) of AMBER, which 
covers the J, H and K bands with $R=30$, and adopted a 50~ms exposure 
time. AMBER unfortunately has poor J-band sensitivity, and we detected
no fringes in that band. The H-band fringes, on the other hand, probe 
significantly higher spatial frequencies than the K-band VINCI would 
have on the same baselines. We mostly chose to not use the FINITO 
fringe tracker \citep{LeBouquin2008}, since our targets were 
at best close to the H=3 limiting magnitude of FINITO. The 
fringes would thus not have been sufficiently stabilized to
allow much longer integration times and compensate the 80\% FINITO 
levy on the H-band flux. \\
Since accurate absolute visibility calibration with AMBER had not 
been demonstrated when we planned the observations, we requested
a very conservative observing strategy. Observations of up to four 
different amplitude calibrators were interleaved with those of each 
science target, to closely monitor the instrumental and atmospheric 
transfer function. Those calibrators were chosen from 
\citet{Merand2004} for angular proximity, well constrained predicted
visibilities, and an approximate magnitude and color match to the 
science targets. The last requirement had to be relaxed somewhat
for the M dwarf targets, since we could locate no apropriate M-type 
calibrators. For those stars we thus used K giant calibrators, which
remain fairly close in near-IR colors. Since AMBER had never been
used to measure precise angular radii, we chose to reobserve
two stars previously measured with VINCI, GJ\,551 and GJ\,887. VINCI has a well established record of accurate
amplitude calibration, and those three stars provide a valuable
check on any potential systematics in the AMBER measurements.

GJ\,879 was observed as a backup target in very poor atmospheric 
conditions (seeing FWHM $>$ 1.5" and $\tau_0$ $<$ 2ms), and has
the smallest angular radius in our sample (highest $V^2$).
Its radius measurement, as a consequence, has significantly
larger error bars than that of any other AMBER source. 

\subsection{Data reduction}

With angular sizes under 2~mas, our targets are only partially resolved
on the longest baselines available for our observations (128 m on the
A0-K0 baseline with AMBER, and 140m on the B3-M0 baseline with VINCI). 
Their squared visibilites $V^{2}$ remain above 0.5 in the H-band. We therefore
cannot derive their angular diameters from just the location of the first
null of the visibility function, and instead need accurate calibration 
of the visibilities. The very partial resolution, on the other hand,
ensures that bandwidth-smearing effects are negligible. At our precision
level on the visibilities, accounting for the finite spectral bandwidth
would only become necessary for $V^2$ below 0.3 for VINCI, and well
beyond the first visibility null for AMBER.

The end products of both the VINCI and the AMBER pipelines consist of 
coherence factors ($\mu^2$) together with an internal error estimate
on that quantity. The coherence factor is related to the squared 
visibility, $V^{2}$, through:

\begin{eqnarray}
V_{\lambda}^{2} = \frac{\mu_{\lambda}^{2}}{T_{\lambda}^{2}},
\end{eqnarray}
where $T^{2}$ is the squared transfer function of the instrument plus
the atmosphere.

Accurate calibration of the absolute squared visibilities therefore critically 
depends on a well understood transfer function. That function is sensitive
to both instrument stability and atmospheric conditions, and can fluctuate
during a night. To assess its stability, we calculated the transfer 
function for every calibrator exposure during a night, disregarding
only those datasets for which too few scans/frames passed the reduction 
pipeline's thresholds to ensure the statistical significance of the 
resulting coherence factor measurement. In most cases, the transfer
function for each target measurement was evaluated from two different 
calibrators, providing some control for temporarily deteriorated 
atmospheric conditions. Using two calibrators also protects against
systematics introduced by a poorly chosen calibrator, such as
unrecognised binaries. When no calibrator was observed immediately 
before or after a target point, we adopted the mean of the transfer 
function for the two nearest calibrators, provided they were observed
within 1h of the science measurement. A few observations had to
be discarded because no sufficiently close pair of calibrator observations
was available. Table \ref{tab-3} and \ref{tab-2}  respectively summarise 
the calibrators' properties for the AMBER and the VINCI observations.

The contribution of the calibration to the visibility error bars accounts
for both the statistical uncertainty on the coherence factor and the 
propagation of the uncertainty on the calibrators' diameters. 
Since the different $V^2$ measurements for a given target share any
systematic error on the calibrators' diameters, they cannot 
be considered as fully independent. Those correlations are 
accounted for in the error bars, using the method described by 
\citet{Perrin2003}. That method is directly applicable to 
VINCI observations, but needed adaptation for AMBER, as we 
detail later.

\subsubsection{VINCI data reduction}
We used V. 3.1 of the standard VINCI reduction pipeline 
\citep{KervellaSegransan2004}. We used the wavelet spectral density 
\citep{Segransan99} as a visibility estimator, as we found that it 
more robustly removes pistonned interferograms than the more common 
Fourier analysis. The commissionning state of the instrument and of
the VLTI array often affected the observing strategy, and we had
to discard a significant number of measurements for which no 
calibrator observations where recorded at the same scan frequency.
\citet{KervellaSegransan2004} showed, from observations of brighter
calibrators, that the squared transfer function of VINCI instrument 
is stable to within 1.5\% during a night. We could therefore 
average the transfer function measurements from all calibrators 
observed during one night, and the observed dispersion mainly reflects 
the statistical errors on these individual measurements. Those are the
main source of uncertainties on our final radii obtained with VINCI.

\subsubsection{AMBER data reduction}

\paragraph{General description}
We used the AMBER data reduction pipeline described by \citet{Tatulli2007}. 
After data reduction, we noted structures in the $V^2$ data obviously 
resulting from correlations in the dataset presumably due to 
non-optimal pipeline settings. Indeed, in february 2008, the AMBER 
TaskForce team \citep{ATF2008} made a report of the displacement of the 
photometric channels with respect to the interferometric channel between 
2004 and 2008. Correlations occur on several pixels if those displacements 
are not correctly calibrated, a step that is achieved through the AMBER 
standard calibration matrix computation. During our October 2007 observing 
run, the spectrograph entry slit was tilted and badly focused, thus 
propagating correlations over 4 to 5 pixels on the detector while 
AMBER spectral resolution is usually sampled over 2 pixels. We corrected this effect
by taking into account integer pixel channel offsets during spectral 
reshifting procedure, to avoid additional correlations to appear during 
the subtraction of the bad-pixel map at the sub-pixel level.
None of our targets had sufficient J-band flux to offset the poor transmission
of AMBER in that band. We therefore discarded the J-band data
as well as wavelengths affected by telluric absorption, only keeping
the centers of the H and K bands, 1.65 - 1.85 $\mu$m and 2.10 - 2.40 $\mu$m.

We concatenated all observations of each object (usually 5 sets
of 1000 frames) into a single dataset from which we could more 
robustly select on fringe SNR, to reject pistonned interferograms 
as well as scans with low flux on one of the two telescopes in
a pair. We verified that selecting the best 20\% to 75\% frames 
produced similar results, and therefore that the details of
the selection do not unduly affect the outcome. Our final
was to keep the best 20\%, as a trade-off between fringe jitter 
suppression and increased noise. Selection on an absolute SNR 
threshold would more effectively reject blurred fringes (which
bias down $V^2$) than accepting a specified fraction of the data, 
but that alternate mode is not available in the current
version of the AMBER pipeline. Another limitation of that
data reduction package is that it uses a common threshold
for all 3 baselines, in spite of their quite different 
throughputs. Those limitations would become more critical 
for datasets containing smaller number of scans than we used
here.

\paragraph{Transfer function}

\begin{figure}
  \centering 
    \includegraphics[width=8cm]{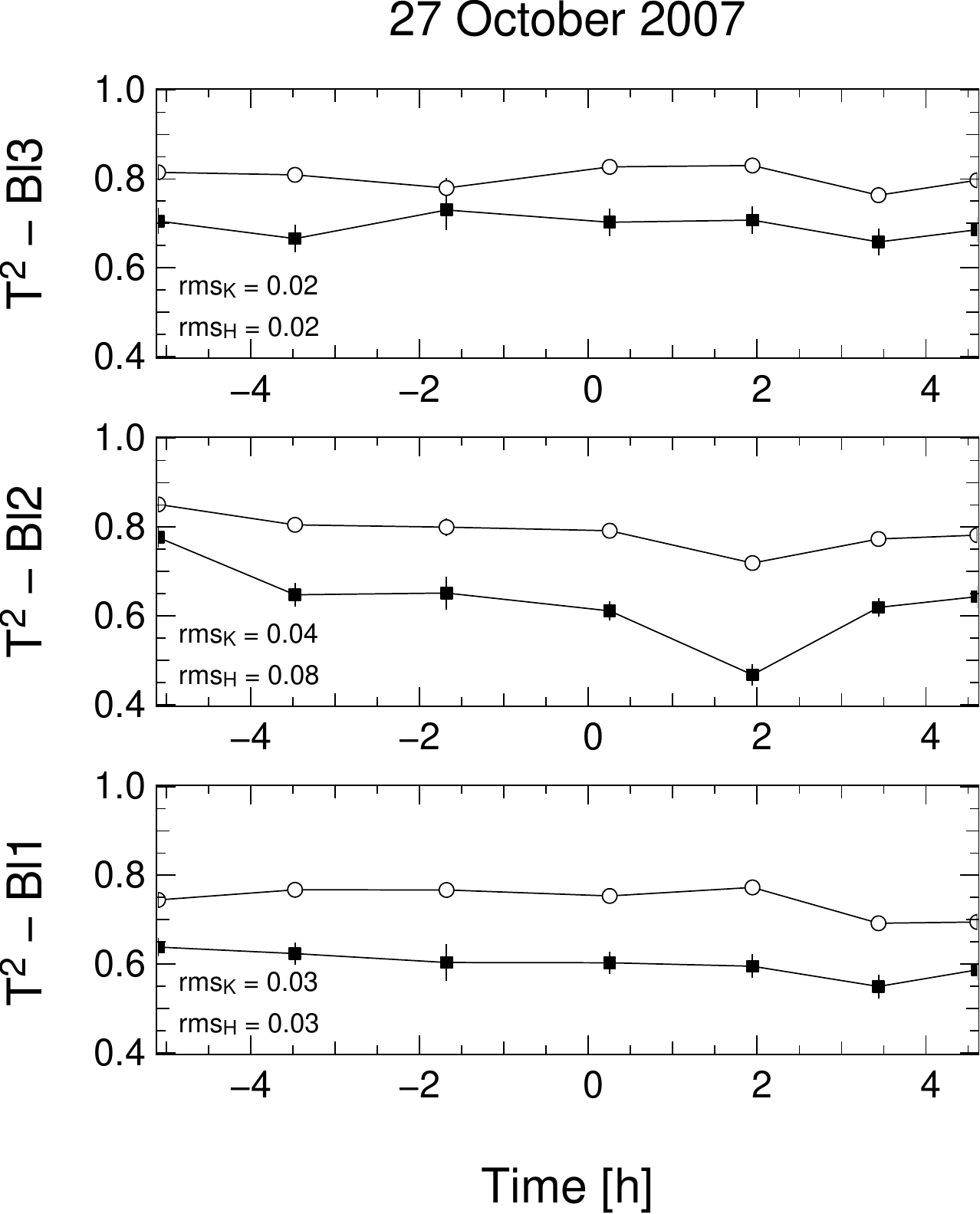}
   \caption{AMBER squared transfer function for Oct 27, 2007. The upper, 
    middle and bottom panel respectively represent baselines 3, 2 and 1. 
    Circles and squares respectively represent the values for the center
    wavelengths of the K-band (2.25$\mu$m) and of the H-band (1.76$\mu$m),
    The corresponding rms dispersions are reported in each panel.}
   \label{Fig-T1}
    \end{figure}

As discussed above, a well understood transfer function is critical 
to measuring absolute visibilities. The stability of the VINCI transfer
function is well established, and a wealth of absolute visibilities 
have been published with that instrument. The reliability of AMBER for 
absolute $V^2$ measurements, on the other hand, has been questioned
and the stability of the AMBER $T^2$ needs to be established. 
Figure~\ref{Fig-T1} shows the squared transfer functions of the 3
AMBER baselines during the night of 27 October 2007. The atmospheric 
conditions during the first half of the night were representative of 
Paranal, with a 0.8 arcsec seeing and a coherence time slightly above 3~ms.
With a 2 to 4\% rms dispersion for both bands on baselines 1 and 3 (90 and 
128 m. length respectively) and in K-band for baseline 2 (90 m.), AMBER approaches the stability of the
VINCI transfer function.
The end of the night had severely degraded atmospheric conditions 
(2.5~arcsec seeing and 1~ms coherence time), and the transfer function
degraded only moderately, except on baseline 2, that had shown a 13\% rms dispersion.

\paragraph{Error bars computation}

The reliability of absolute $V^2$ measured with AMBER has not
yet been well established, and evaluating realistic error bars 
therefore needs close attention. The 16 to 18 spectral channels
which we usually kept (7 to 8 in the H-band, and 9 to 10 in the 
K-band) are measured simultaneously. They therefore share the 
same atmosphere, as well as any error on the angular diameter
of the calibrator(s). If those factors dominate over statistical
noise, the individual channels become highly non-independent.

To quantify theses correlations between spectral channels, 
we generalise the formalism developed by \citet{Perrin2003} 
for a two-channel combiner. For two Gaussian distributions of 
$V^2$ series, the correlation coefficient between spectral channel 
$k$ and a reference channel, $r$, is:

\begin{eqnarray}
\rho_{\lambda_{r},\lambda_{k}} = \frac{\langle(V_{\lambda_{r}}^{2}-\overline{V_{\lambda_{r}}^{2}}) (V_{\lambda_{k}}^{2}-\overline{V_{\lambda_{k}}^{2}})\rangle}{\sqrt{ (V_{\lambda_{r}}^{2}-\overline{V_{\lambda_{r}}^{2}})^2 (V_{\lambda_{k}}^{2}-\overline{V_{\lambda_{k}}^{2}})^2}},
\end{eqnarray}

\noindent we computed $\rho_{\lambda_{r},\lambda_{k}}$ through Monte-Carlo simulations from the mean and variance of $V_{\lambda_{r}}^{2}$ and $V_{\lambda_{k}}^{2}$.
This provides a global correlation factor for a given band and to 
estimate amplitude of the error bar that is independent, and then 
to compute realistic error on the final diameter. As expected, 
we found that $V^2$ data in a same spectral band are highly correlated 
and thus biases the final result if correlations are not taken into account.

Error estimates on final radii obtained with AMBER mainly come from 
transfer function uncertainties and correlations between squared visibilities. 
Those factors take more importance as the coherence time degrades.

\subsection{Data analysis}

\subsubsection{Limb-darkened diameters}

At the level of accuracy achieved on the diameter determination for K dwarfs, discrepancy between uniform disk (UD) and limb-darkened disk (LD) is significant. Therefore, we used the non-linear limb-darkening law describing the intensity distribution of the star disk from \citet{Claret2000} :

\begin{equation}
I(\mu) = I(1)\left[1- \sum_{k=1}^4 a_{k}(1-\mu^{k/2})   \right],
\end{equation}

\noindent where $I(1)$ is the specific intensity at the center of the disk, $\mu = cos \gamma$,  $\gamma$ being the angle between the line of sight and the emergent intensity, and $a_{k}$ the limb-darkening coefficients. $T_{eff}$ and $log$ g for each target are shown in table  \ref{tab-6}, with the corresponding references. We used those parameters, added to the photometric band (K for VINCI, H+K for AMBER) and micro-turbulence velocity (assumed to be $V_{T}=$ 2km/s), to select the corresponding limb-darkening coefficients for the PHOENIX models.

Then, we adjusted a limb-darkened disk that is given by \citet{Davis2000} to the $V^2$ data :

\begin{equation}
V_{LD, \lambda} = \frac{\int_{0}^{1}{ d\mu I\left(\mu \right) \mu J_{0}\left(\pi B \theta_{LD}/\lambda \left(1-\mu^{2}\right)^{1/2}\right)  }}{\int_{0}^{1}{ d\mu I\left(\mu \right)}}
\end{equation}

\noindent We used the same LD coefficients for H and K bands since corresponding discrepancy has been evaluated at the 0.05\% level on the final radius, negligible as compared to the error bar amplitude obtained with AMBER. Uncertainties on the same coefficients due to a slightly different $T_{eff}$ and $log$ g can also be neglected, a 200K change corresponding to a 0.01\% on final radius estimate.

\begin{figure}
  \centering 
    \includegraphics[width=8cm]{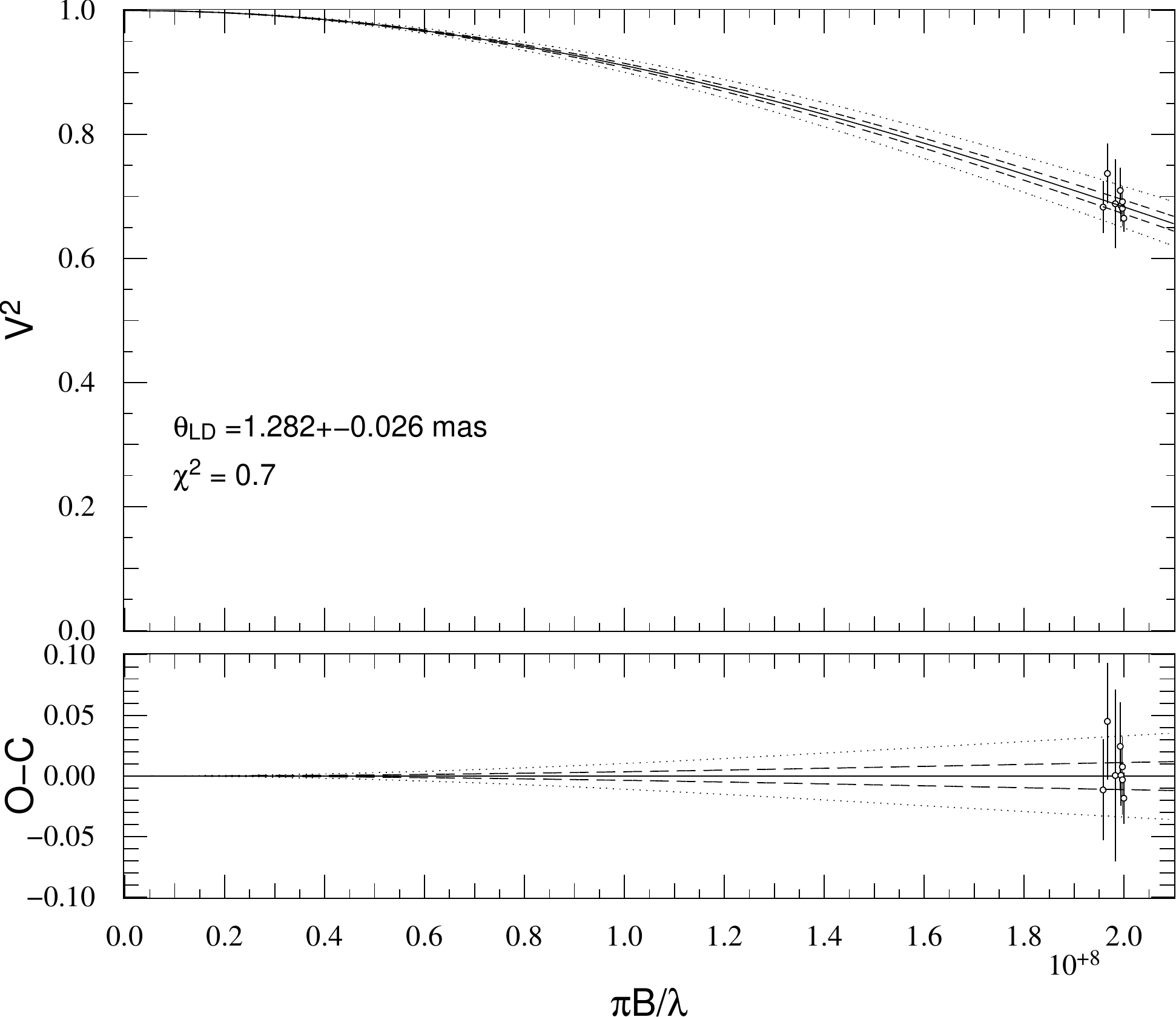}
   \caption{Calibrated squared visibilities from VINCI and
            best-fit LD disk model (solid) for GJ663A vs. spatial frequencies.  1-$\sigma$ (dash) and 3-$\sigma$ (dot) uncertainties are also indicated.
            }
   \label{Fig-1}
    \end{figure}

\begin{figure}
  \centering 
    \includegraphics[width=8cm]{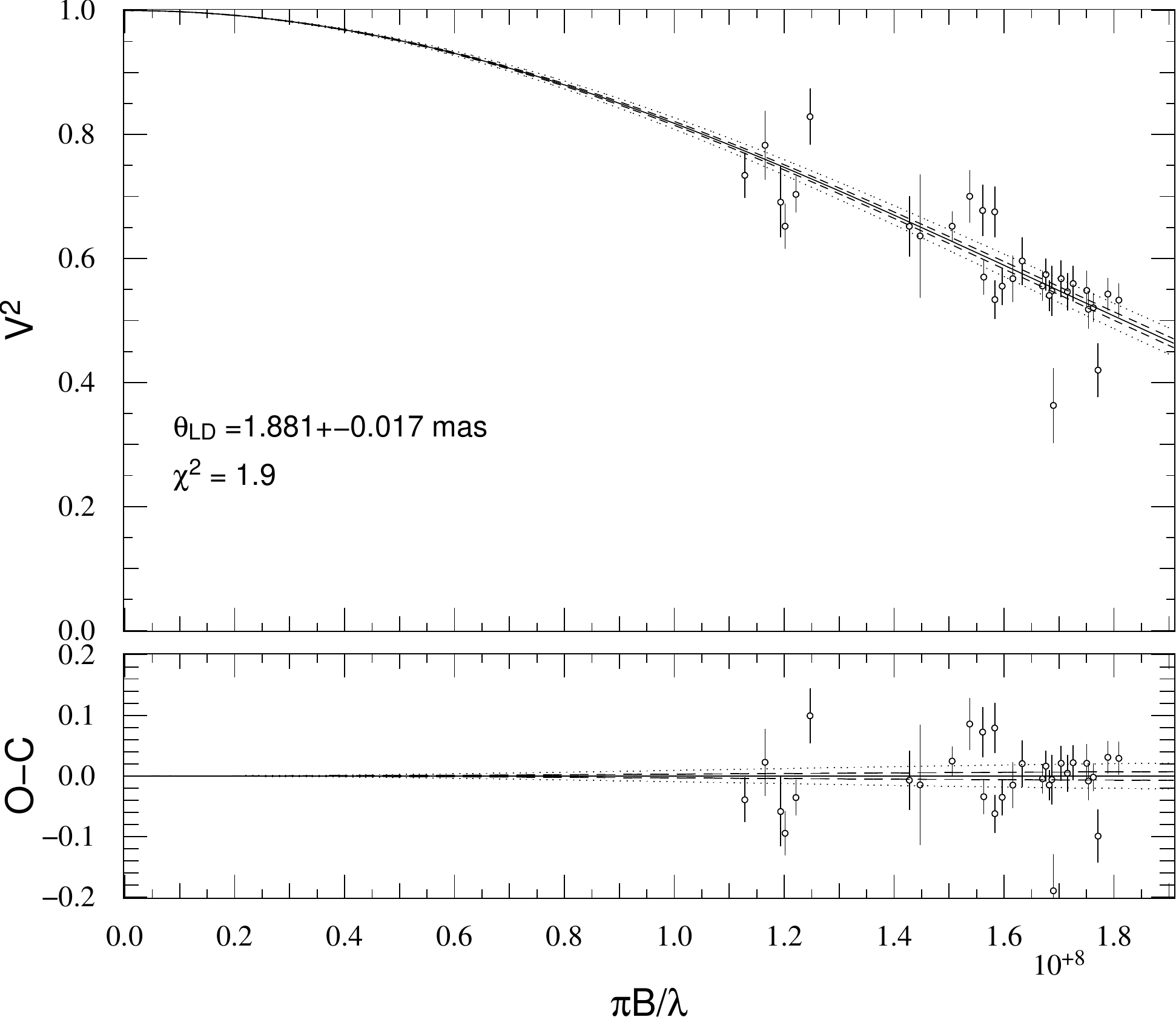}
   \caption{Calibrated squared visibilities from VINCI and
            best-fit LD disk model (solid) for GJ845 vs. spatial frequencies.  1-$\sigma$ (dash) and 3-$\sigma$ (dot) uncertainties are also indicated.
            }
   \label{Fig-3}
    \end{figure}

\begin{figure}
  \centering 
    \includegraphics[width=8cm]{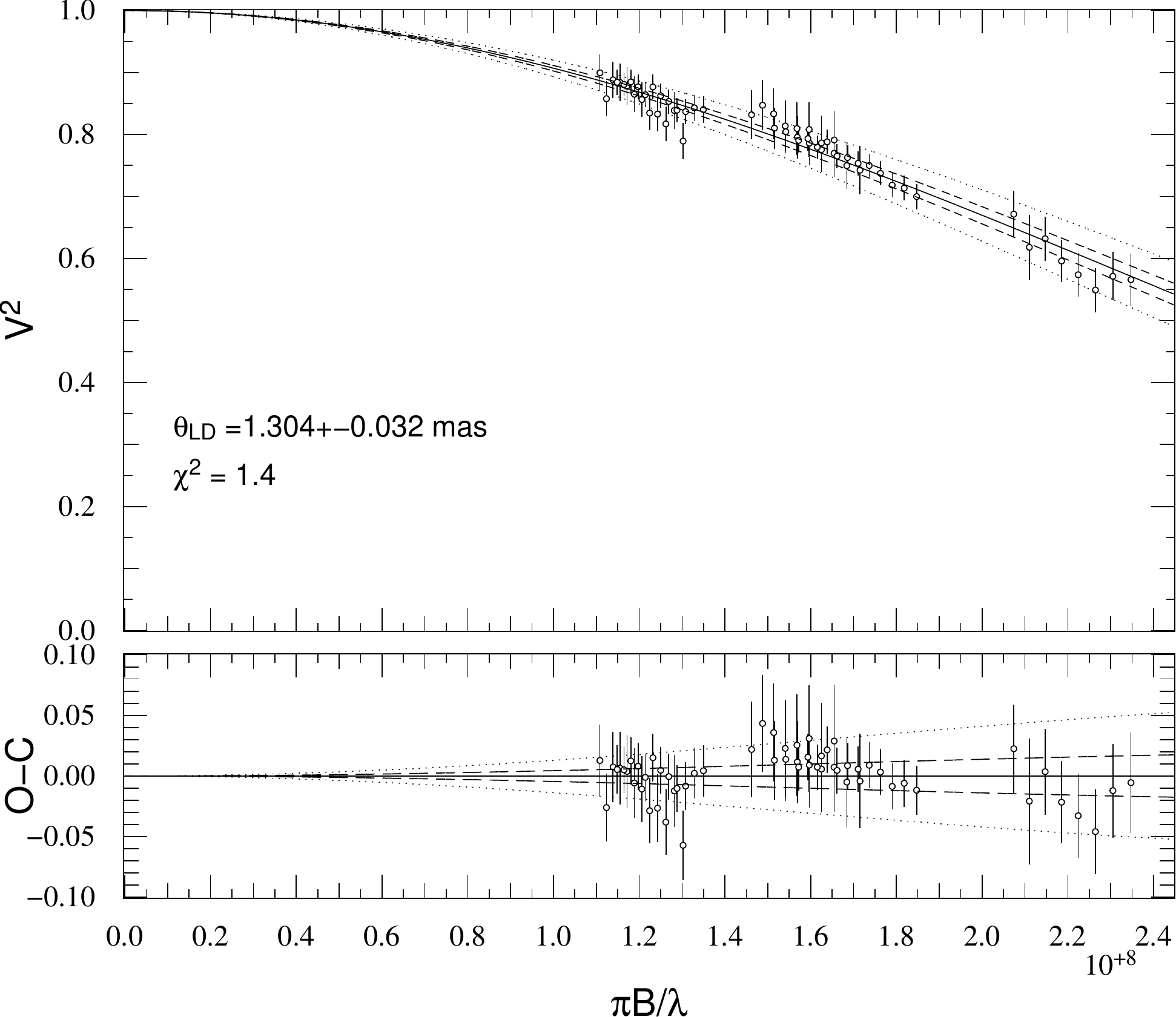}
   \caption{Calibrated squared visibilities from AMBER (low-resolution mode) and
            best-fit LD disk model (solid) for GJ887 vs. spatial frequencies.  1-$\sigma$ (dash) and 3-$\sigma$ (dot) uncertainties are also indicated. Error bars amplitudes include both correlated and non-correlated errors.
            }
   \label{Fig-5}
    \end{figure}

\begin{figure}
  \centering 
    \includegraphics[width=8cm]{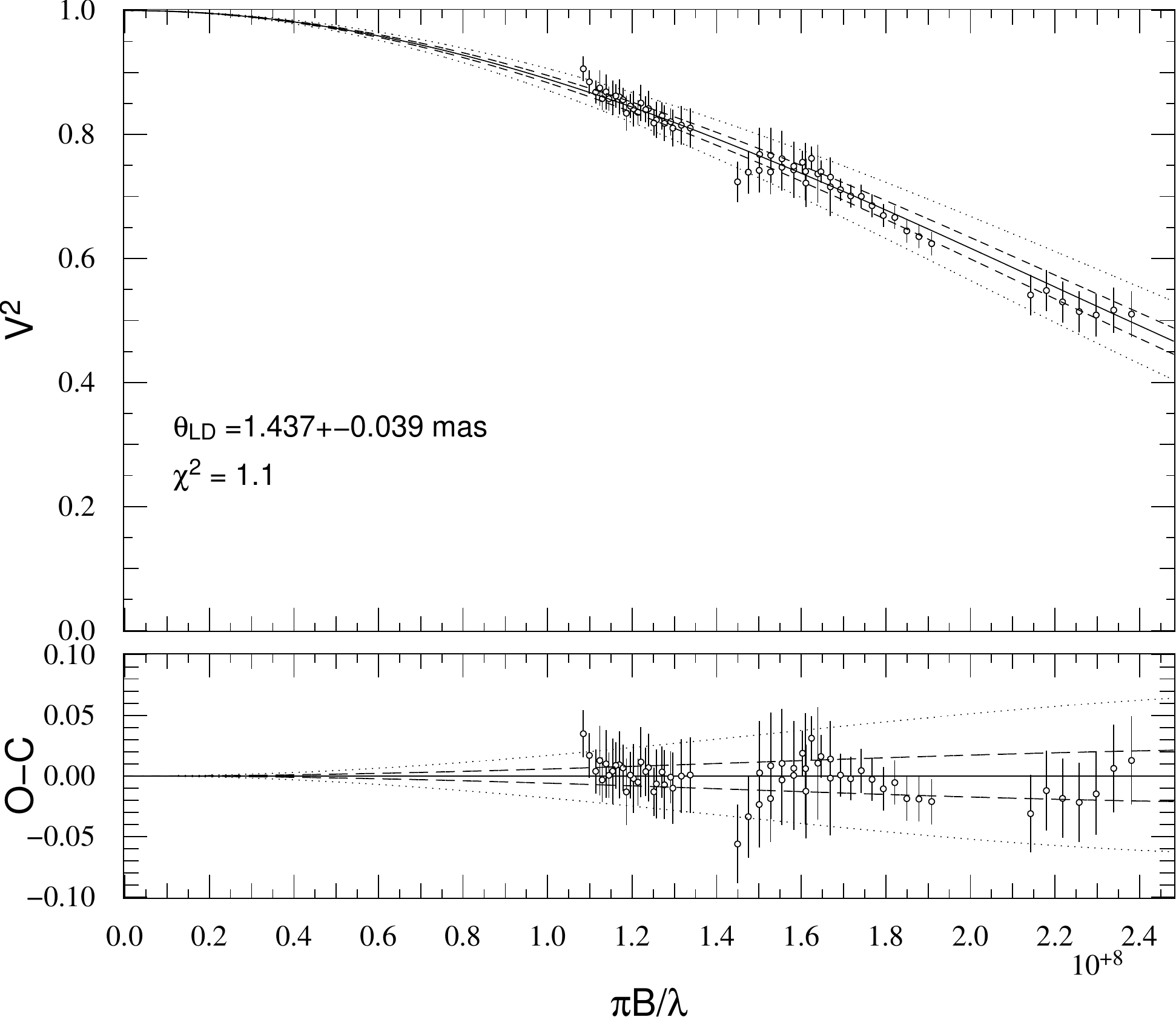}
   \caption{Calibrated squared visibilities from AMBER (low-resolution mode) and
            best-fit LD disk model (solid) for GJ166A vs. spatial frequencies.  1-$\sigma$ (dash) and 3-$\sigma$ (dot) uncertainties are also indicated. Error bars amplitudes include both correlated and non-correlated errors.
            }
   \label{Fig-6}
    \end{figure}

\noindent  Table \ref{tab-5} lists the derived UD and LD angular diameters and their corresponding errors for both instruments.
Figures \ref{Fig-1}, \ref{Fig-3} and \ref{Fig-2} show $V^2$ data from VINCI with corresponding fitted LD models for GJ\,663\,A, GJ\,845 and GJ\,570\,A respectively.
Figures \ref{Fig-5}, \ref{Fig-6}, \ref{Fig-7} and \ref{Fig-8} show $V^2$ data from AMBER with corresponding fitted LD models for GJ\,887, GJ\,166\,A, GJ\,551 and GJ\,879 respectively.
   
\subsubsection{Instrument systematics}

Assessing systematics is essential to consider when reaching a few percent precision on angular diameters. We thus wanted to check for consistency between both instruments on GJ\,887 and GJ\,551. The results we obtain are shown in Table \ref{tab-5}. Angular diameter determinations are consistent for both instruments. GJ\,887 has been observed in optimal conditions in both cases and the agreement is good at  1-$\sigma$ level, as it is for GJ\,551.
Bracketing each source point with two calibrators allow a better sampling of the transfer function, thus slightly reducing those effects as shown on fig. \ref{Fig-T1}. When good atmospheric conditions are met and a proper calibration applied, systematics on H and K band can be expected to be at 2\% level, slightly above VINCI's.

\section{Discussion}
We focussed our study on the low and very low end of the main sequence with spectral types ranging from  K0.5 to M5.5. The stars composing our sample 
cover a wide range of masses - from 0.12 to 0.8~M$_{\odot}$ - which results in different physical conditions affecting their internal structure, 
the heat transport, as well as their evolution. Their atmosphere chemistry is also strongly affected with the disappearance of true continuum to the benefit of
a complex and a high gravity stellar atmosphere with strong molecular absorptions bands. Metallicity and activity also play a role, although we expect it to 
be a second order effect, at least in the near-infrared. In the following sections, we discuss the implications and constraints brought by our measurements on stellar physics modelling.

\subsection{Luminosity-radius relationship}

\begin{figure}
  \centering 
\includegraphics[width=8cm]{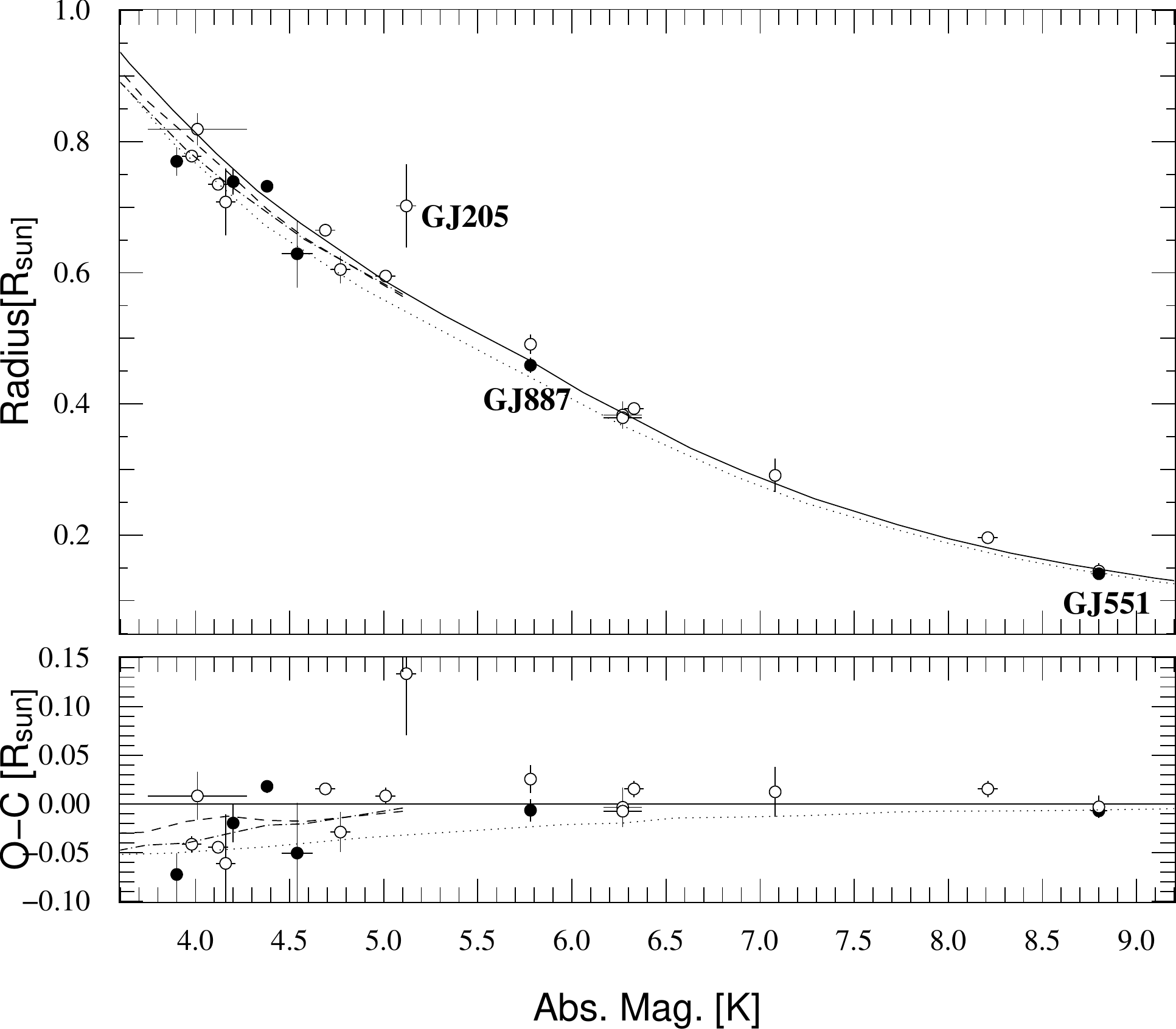}
\caption{Luminosity-Radius relationship - Single stars radii vs. absolute K magnitudes superimposed on 5 Gyr isochrones theoretical models \citep{Baraffe1998}. Our results from AMBER and VINCI are shown as filled circles. We have also included other stellar radii from the litterature determined by interferometry : \citet{Berger2006} for GJ\,15\,A, \citet{Boyajian2008}, \citet{diFolco2007}, \citet{Kervella2008}, \citet{Segransan2003} and \citet{Lane2001} as empty circles. Only radii measurements better than 10\% are displayed. Different models for 5 Gyr isochrones are also shown : solar metallicity with $L_{mix} = 1.0 H_P$ (solid), $L_{mix} = 1.5 H_P$ (dash) and $L_{mix} = 1.9 H_P$ (dashdot) as well as a metal deficient, [M/H]=-0.5 model with $L_{mix} = 1.0 H_P$ (dot). GJ\,205, GJ\,887 and GJ\,551 that appear in the discussion are labeled.}
\label{Fig-9}
\end{figure}

We have first chosen to compare our results to  \citet{Baraffe1998}  models, in a luminosity-radius diagram which corresponds to the observables. Indeed, it has the advantage of avoiding the inclusion of the mass-luminosity (hereafter ML) relationship, for which reliability regarding K dwarfs has not been demonstrated yet. Figure \ref{Fig-9} shows our VLTI results.  Different sets of models are overplotted, ie.  for an age of 5 Gyr, featuring different mixing lengths \citep{Bohm1958}, $L_{mix}$ expressed in pressure scale height $H_P$, that allow to assess the convective efficiency, as well as two distinct metallicities : $[M/H]=0$ and $[M/H]=-0.5$.

 \citet{Baraffe1998}  models are  in excellent agreement with our observations in the very low-mass part of the luminosity-radius diagram. The radius determined for \textit{Proxima} (GJ\,551) is perfectly reproduced by theory. In this part of the relation, stars are fully convective which greatly simplifies the modelling of heat transport and therefore, our result validates the equation of state used by \citet{Baraffe1998}.
GJ\,887 is an early type M dwarf, located slightly above the boundary of this class of objects. The radius determined for GJ\,887 is also in perfect agreement with model predictions. Other measurements from the literature confirms that stellar interior physics for this mass range are well mastered.

At the time of writing this paper, there are relatively few radii measurements that would allow a discussion in the 0.5 - 0.75 M$_\odot$ region. 61 Cyg A and B radii have been recently determined by \citet{Kervella2008} and are also well reproduced by theory. We note that \citet{Berger2006} published 6 radii measured with the CHARA array in this part of the diagram. The authors claim discrepancies with models at the 2 to 3 $\sigma$ level.  Such large departures from theory have not been observed by other studies. One may note, however, that 5 of the 6 stars measured by \citet{Berger2006}  have inflated radii. Those  stars were measured with the instrument "CHARA-Classic", a recombiner that does not include a single-mode filtering. Such measurements are prone to systematic calibration errors and indeed, the one star (GJ\,15\,A) which they measured with "CHARA-FLUOR" (instrument with single mode filtering)  is in excellent agreement with the models. Although a possible explanation, we note that such instrumental effect is expected to result in a uniform dispersion. We decided, however, not to include those results in this discussion.

GJ\,205 radius, as measured with VINCI, is about 15\% above models. Radial-velocity measurements on this object have not revealed any massive (heavier than a Saturn-mass) companion (X. Bonfils, priv. comm.) that would have induced a lower interferometric visibility, thus a larger radius. Moreover, GJ\,205 has not been reported to show significant activity \citep{LopezMorales2007}. Nevertheless, this star is more luminous than other known objects belonging to the same spectral class and is probably inflated.

In the upper part  of the luminosity-radius relationship, models reproduces the observations provided that larger  mixing length are used (such as $L_{mix} = 1.5 H_P$ and $L_{mix} = 1.9 H_P$). This part of the relationship may be used to calibrate $L_{mix}$ provided accurate observational radii and magnitude determination are available.

\begin{figure}
  \centering 
\includegraphics[width=8cm]{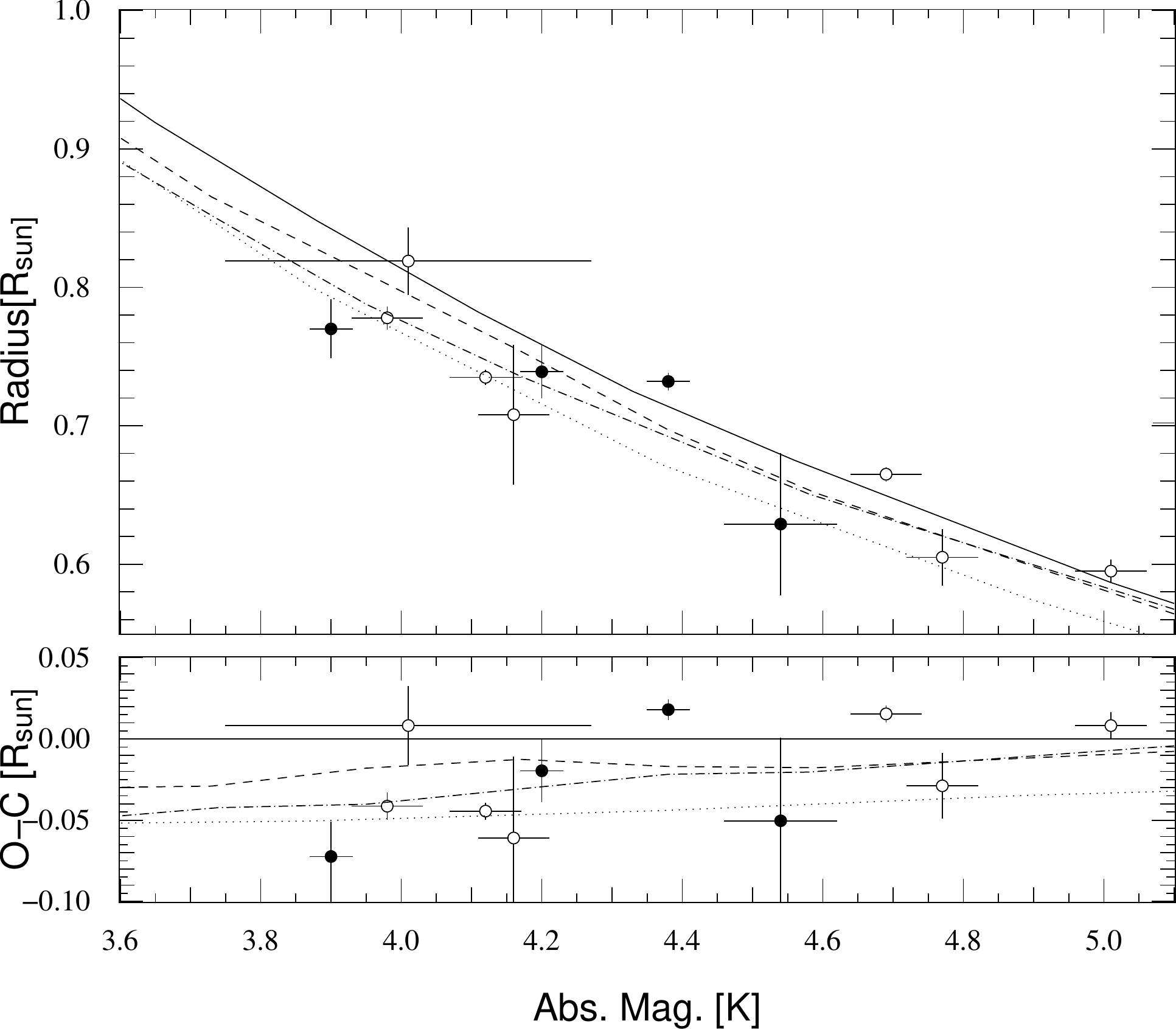}
\caption{Luminosity-Radius relationship - same as fig. \ref{Fig-9}, zoomed on the upper part of the diagram. }
\label{Fig-9b}
\end{figure}

Figure \ref{Fig-9b} displays a zoom on this area of the relation. 

Unfortunately, GJ\,663\,A does not appear in the tables nor in the graphs because of the lack of K magnitude measurements. Its measured radius is however shown in table \ref{tab-5} for completeness. It should be noted that some efforts are needed to obtain accurate near-infrared photometry of nearby K dwarfs to tighten the mass-luminosity relation and therefore better constrain theoretical models in the upper part of the luminosity-radius relationship.

\subsection{Mass-radius relationship}

\begin{figure*}
  \centering 
\includegraphics[width=17cm]{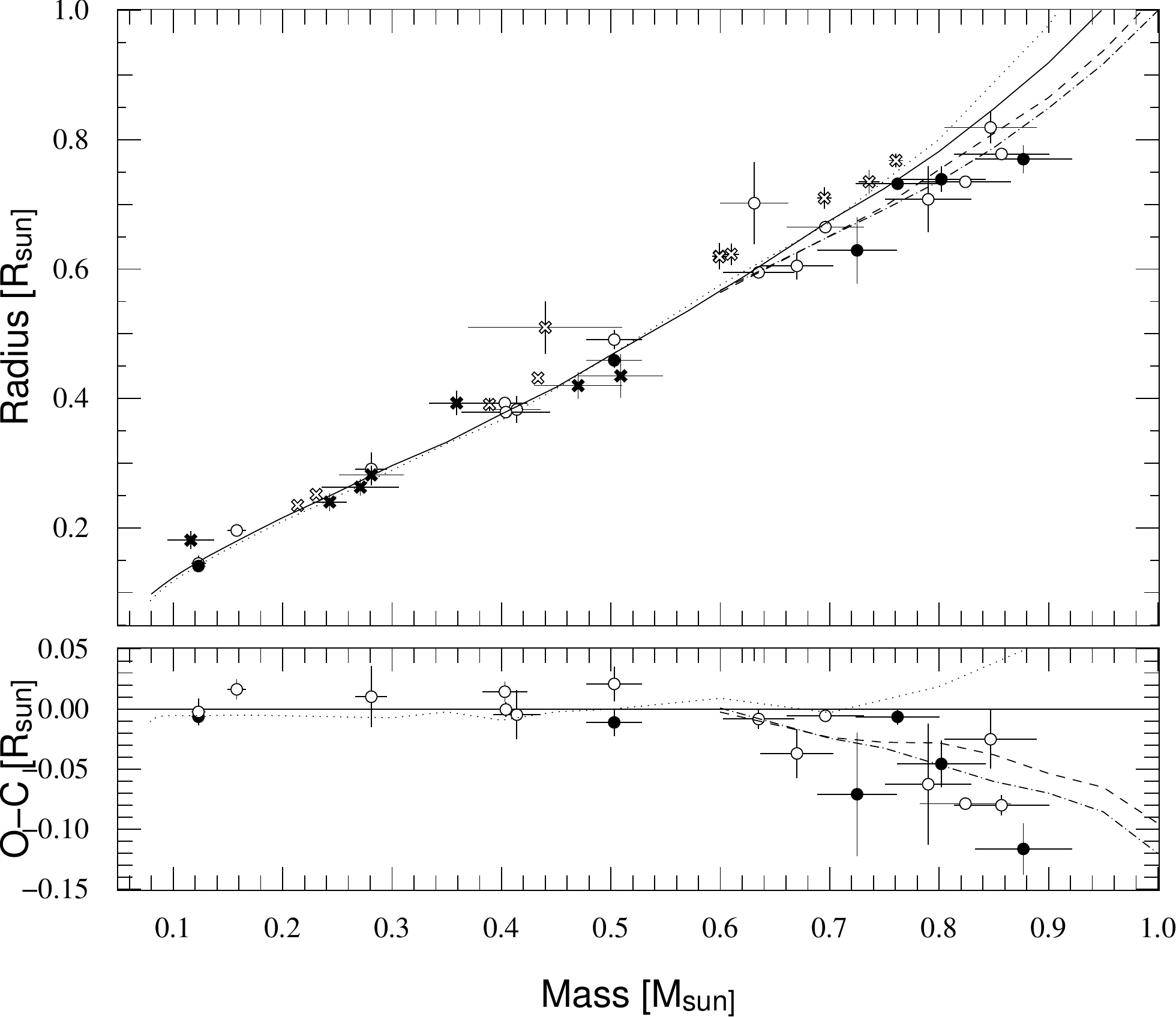}
\caption{Mass-radius relationship - masses and radii superimposed on 5 Gyr isochrones theoretical models \citep{Baraffe1998}. Our results appear as filled circles. Other long baseline interferometry measurements come from PTI \citep{Lane2001}, VLTI \citep{Segransan2003} and CHARA-FLUOR: \citet{Boyajian2008}, \citet{diFolco2007}, \citet{Kervella2008} and \citet{Berger2006} for GJ15A, all as empty circles. Solar metallicity with $L_{mix} = 1.0 H_P$ (solid), $L_{mix} = 1.5 H_P$ (dash) and $L_{mix} = 1.9 H_P$ (dashdot) are shown as well as a metal deficient, [M/H]=-0.5 model with $L_{mix} = 1.0 H_P$ (dot).   Only radii measurements better than 10\% are displayed. Solar neighboorhood eclipsing binary measurements are represented as empty crosses while OGLE-T transiting binaries are represented in filled crosses. 
Only residuals from long-baseline interferometry results are displayed.}
\label{Fig-10}
\end{figure*}

The translation of our direct measurements into a mass-radius diagram requires the use of an empirical ML relationship. We used the relation determined by \citet{Delfosse2000} to compute masses for GJ\,887 and GJ\,551, and a recent one, by \citet{Xia2008} for the 0.7 to 1.0 M$_\odot$ range. This latter study is based on \citet{Henry1993}. The ML relationship for stars below 0.6 M$_\odot$ is built on a large number of accurate masses and luminosities \citep{Segransan2000} and exhibits a very low dispersion in near-infrared. In this part of the mass-luminosity relationship, models reproduce the observations fairly well indicating that both atmosphere and interior physics of very low mass stars are well mastered.
The empirical ML relationship  above 0.6 M$_\odot$ shows a much larger dispersion than for M dwarfs which is  related to the modest accuracy of the masses in this mass  range. The derivation of  masses from absolute magnitude is therefore less accurate than for the lower part of the relation and we adopted an arbitrary error of 5\% on masses determination. Those results appear in Table \ref{tab-5}.
Figure \ref{Fig-10} shows our results in a mass-radius (MR) diagram with results from other studies. Five Gyr model isochrones from \citet{Baraffe1998} are represented for different mixing lengths and stellar metallicity.


\subsection{Stellar properties}

\subsubsection{Effective temperature}

Angular diameter measurements associated with accurate parallax and Johnson photometry allow to derive the star effective temperature. UV to near IR photometry have been obtained from the literature, mostly from \citet{Morel1978}.  Visual and near-infrared photometry appear in table \ref{phot_table}.
We derived effective temperatures, by  inverting the surface-brightness empirical relations calibrated by \citep{Kervella2004}. 
 $T_{eff}$ values determined through this method are in good agreement with $T_{eff}$ determined by spectroscopy, both appear in table \ref{tab-5}.

\subsubsection{Metallicity}
\begin{figure}
  \centering 
\includegraphics[width=8cm]{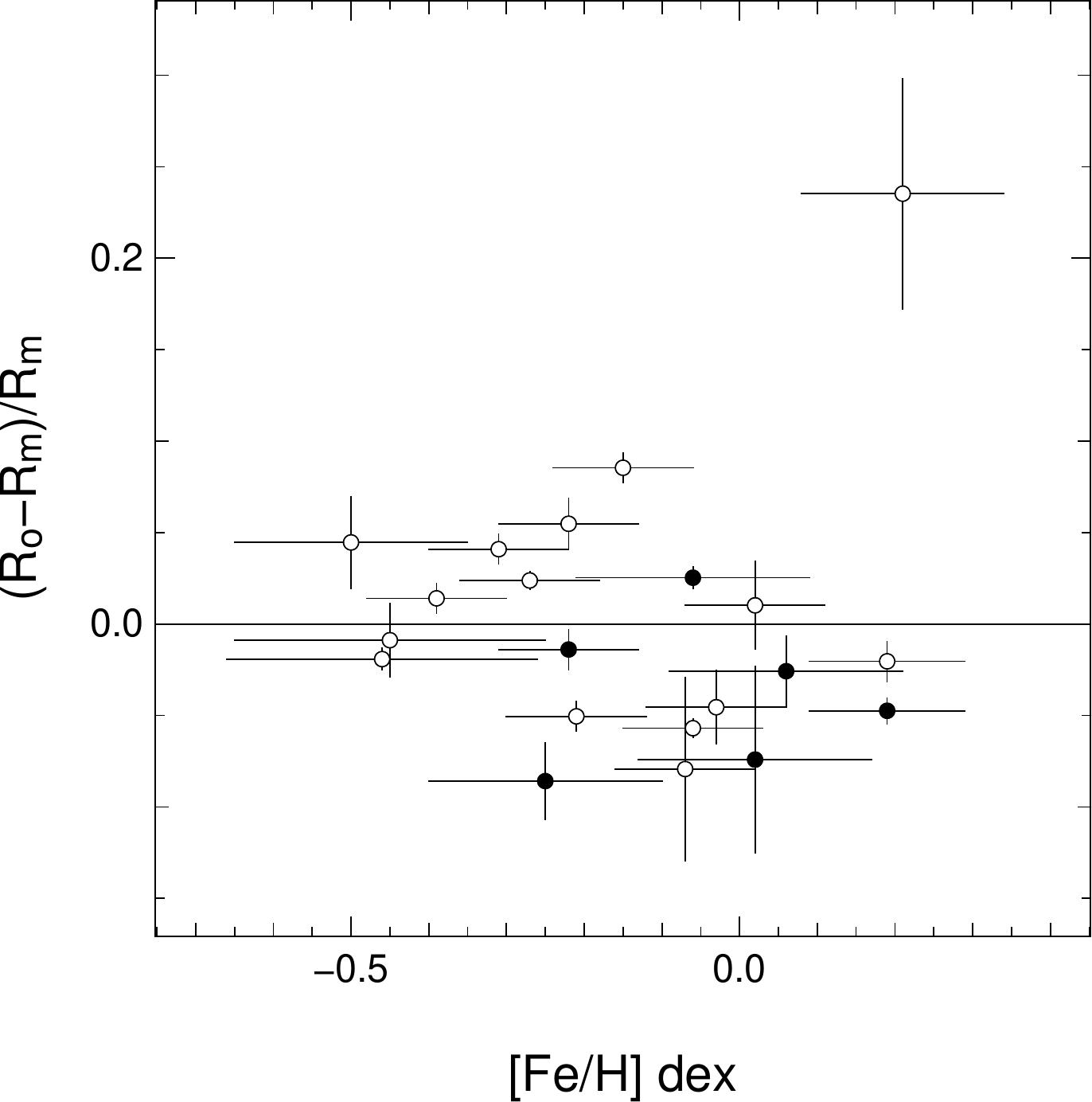}
\caption{Fractional deviation of single stars radii derived by interferometry from a 5-Gyr, $L_{mix} = 1.0 H_P$ model \citep{Baraffe1998}, vs. stellar metallicity.}
\label{Fig-10b}
\end{figure}

Metallicity effects have been mentioned  by \citet{Berger2006}  to explain the difference between four of his measurements and solar metallicity models.
Indeed, the authors claim that missing opacity sources in the models, such as TiO, would explain the models underestimation of stellar radii for some M dwarfs.
\citet{LopezMorales2007} recently  studied the correlation between magnetic activity, metallicity and low-mass stars radii. Based on \citet{Berger2006}
 measurements, she reach the same conclusion. However, no other instrument (PTI, VINCI, CHARA-FLUOR or AMBER) could confirm this hypothesis
 except for  GJ\,205 (VINCI) as explained in Sect. 3.1. Without the "CHARA-Classic" measurements made by  \citet{Berger2006}, the metallicity-radius diagram for single 
 stars (Fig. \ref{Fig-10b}) no longer shows such correlation. 

\subsubsection{Activity}
Close-in eclipsing binaries (EB), for which accurate masses and radii have been measured by several authors such as \citet{Torres2002}, show significant discrepancies with stellar models (see e.g. \citet{Ribas2008}).
Recently, \citet{Chabrier2007} explained  those discrepancies by invoking the reduced convective efficiency and starspots coverage of eclipsing binaries.
The difference is only observed in mass-radius diagrams (and not in the luminosity-radius diagram) because the slightly lower effective temperature of EB is compensated by a larger radius, only slightly changing the luminosity. This explanation is only meaningful for EB with periods of a few days implying heavy 
tidal effects, orbital synchronization, and therefore an enhanced activity. This trend is shown on Fig. \ref{Fig-10}, where EB are represented as empty crosses.

However, the same arguments cannot be used in the case of single stars. Rotational velocity is an excellent hint of stellar activity. While low-mass and very low-mass EB are routinely characterised by $v$sin$i$ between 7.11 km/s (CU Cnc B, \citet{Ribas2003} and 129.5 km/s (OGLE BW3 V38A, \citet{Maceroni2004}, single stars rarely exceed 3 km/s. We assessed stellar activity for our targets thanks to CORALIE spectra and it indeed appears that all of them do not show an activity comparable with EB ones. Corresponding log $R'_{HK}$ and vsin$i$  appear in Table \ref{tab-5}. Furthermore, activity cannot explain radii discrepancies reported by \citet{Berger2006} since none of those single stars belonging to their sample show high activity levels \citep{LopezMorales2007}.

Activity cannot be claimed as the source of deviation in the upper part of the mass-radius relationship for single stars. However, radii of single inactive M dwarfs measured by interferometry are in excellent agreement with models from \citet{Baraffe1998}. Thus, discrepancies pointed out by \citet{Torres2002} and \citet{Ribas2003} only concern fast rotating stars, confirming the fact that rotation strongly affects the internal structure of those objects.

\section{Conclusion}

Those new results obtained at the VLTI  with its near-infrared instruments, VINCI and AMBER, allow to better constrain the mass-radius relationship for low and very low mass stars.  We have shown that AMBER is now able of achieving high quality absolute visibility measurements provided that at least 3 calibrator stars are observed and that a careful data reduction and analysis is conducted. Using VINCI as a benchmark, those results are also shown to be reliable, even if the proposed approach requires a time consuming observing strategy. Assessment of potential systematics as well as realistic error bars estimates have been crucial to compare our results with models in a meaningful way.

Models are in good agreement with the observations, confirming a correct understanding of the underlying physics of low and very low mass stars. The very low-mass regime is almost adiabatic and thus constraints the equation of state. A small mixing length of $L_{mix}=1.0H_P$ leads to a progressive underestimation 
of radii for early K dwarfs. As expected,  the lower convection efficiency in K dwarfs require a significantly greater mixing length to reproduce observed radii of low mass stars.

\begin{acknowledgements}
We are very grateful to Guy Perrin, Florentin Millour and Gilles Duvert for their advices about AMBER data reduction as well to Stanislav Stefl and Carla Gil for their work as VLTI night astronomers. 
B.-O.D also would like to thank Fabien Malbet and Guy Perrin for having organized a school devoted to interferometry at Goutelas in 2006. 
Many thanks also to France Allard and Corinne Charbonnel for fruitful discussions about very low-mass stars atmospheres and interior physics.
We thank the anonymous referee for constructive comments on the manuscript.
This work benefits from the support of the \emph{Fonds National Suisse de la Recherche Scientifique}.
This study has made use of amdlib 2.2 developed by the Jean-Marie Mariotti Center, supported by INSU (CNRS and Minist\`ere de la recherche, France).
This research has made use of the SIMBAD database operated by CDS, Strasbourg, France.
This publication makes use of data products from the Two Micron All Sky Survey, which is a joint project of the University of Massachusetts and the Infrared Processing and Analysis Center/California Institute of Technology, funded by the National Aeronautics and Space Administration and the National Science Foundation.
\end{acknowledgements}


\begin{table*}
    \caption[]{Observation log of stars published in this work. AMBER baseline configuration was always A0-K0-G0 (128-90-90)m during our observing runs with this instrument. Some targets were observed during the shared risk observing period (march 2002 - december 2002) of the ESO Very Large Telescope Interferometer (VLTI).}
      \begin{tabular}{llclclcc}
\hline
Star&  Spectral & $m_{K}$ & Instrument & Date &Baseline & DIMM Seeing  & Mean $\tau_{0}$  \\
         &  Type     &                  &                     &           &                 &       ["]        &             [ms]                \\
\hline \noalign{\smallskip}	
GJ\,663\,A  & K0V& -     & VINCI          &13-05-2003 &B3-M0 (140m.)& - & -                \\
                     &        &              &                      &26-05-2003 &B3-M0 (140m.)& 0.7 & 5                \\
                     &        &              &                      &27-05-2003 &B3-M0 (140m.)& 0.5 & 7                \\
                     &        &              &                      &28-05-2003 &B3-M0 (140m.)& 0.8 & 6                \\
                                                                                                                                      
GJ\,166\,A  & K0.5V& 2.41 & AMBER      &27-10-2007&A0-K0-G1         & 1.2 & 3                \\
          
GJ\,570\,A  & K4V & 3.06 & VINCI           &07-04-2003&B3-M0 (140m.)& 0.6 & 4                \\         
                     &          &         &                       &08-04-2003&B3-M0 (140m.)& 0.5 & 6                \\                            
                     &          &         &                       &12-04-2003&B3-M0 (140m.)& 0.7 & 6                \\    
                     &          &         &                       &15-04-2003&B3-M0 (140m.)& 0.5 & 9                \\                         
                     &          &         &                       &16-04-2003&B3-M0 (140m.)& 0.7 & 5                \\    
                     &          &         &                       &21-04-2003&B3-M0 (140m.)& 0.7 & 4                \\                         
                     &          &         &                       &27-04-2003&B3-M0 (140m.)& 0.5 & 3                \\     
                     &          &         &                       &07-05-2003&B3-M0 (140m.)& 1.2 & 2                \\                                              
                     &          &         &                       &08-05-2003&B3-M0 (140m.)& 1.2 & 3                \\     
                     &          &         &                       &09-05-2003&B3-M0 (140m.)& 0.6 & 7                \\          
                                                                                                          
GJ\,845\,A  & K5V & 2.18 & VINCI           &15-09-2002&E0-G1 (66m.)  & 1.1 & 2                \\   
                     &          &         &                       &16-09-2002&E0-G1 (66m.)  & 1.0 & 2                \\                            
                     &          &         &                       &17-09-2002&E0-G1 (66m.)  & 0.8 & 2                \\    
                     &          &         &                       &10-10-2002&B3-M0 (140m.)& 0.9 & 2                \\                         
                     &          &         &                       &12-10-2002&B3-M0 (140m.)& 1.1 & 4                \\    
                     &          &         &                       &16-10-2002&B3-M0 (140m.)& 0.8 & 7                \\                         
                     &          &         &                       &17-10-2002&B3-M0 (140m.)& 1.2 & 5                \\     
                     &          &         &                       &19-10-2002&B3-M0 (140m.)& 0.6 & 4                \\                                              
                     &          &         &                       &22-10-2002&B3-M0 (140m.)& - & -                \\     
                     &          &         &                       &26-10-2002&B3-M0 (140m.)& 0.9 & 3                \\          
                                          
GJ\,879      &  K5Vp& 3.81  & AMBER     &03-10-2008&A0-K0-G1         & 1.4 & 1                \\

GJ\,887    &   M0.5V& 3.36  & AMBER    &27-10-2007&A0-K0-G1         & 1.2 & 3                \\
                                               
GJ\,551    &   M5.5V& 4.38  & AMBER    &27-02-2008&A0-K0-G1         & 1.0 & 3                \\

\hline
         \end{tabular}
         \label{table_obs}
   \end{table*}

\begin{table*}
    \caption[]{List of calibrator stars used during VINCI runs. Angular diameters come from \citet{Borde2002} and  \citet{Merand2004}.}
   
      \begin{tabular}{llclcc}
      \hline
Calibrator     &      Target    & Ang. dist. & Spectral& $m_{K}$ & $\theta_{UD}$ K band  \\
                       &                      & degrees   & Type      &           &   [$mas$]                           \\
\hline \noalign{\smallskip}	

HR\,8685        &   GJ\,845                       & 78.3, 19.3  & M0III & 1.98  & 2.01$\pm$0.02  \\
$\delta$ Phe&    GJ\,845                        & 31.5 &  G9III & 1.63  & 2.19$\pm$0.02  \\
HR\,8898       &   GJ\,845                        & 12.7 & M0III   & 1.81  & 2.31$\pm$0.03   \\
HD\,130157  &  GJ\,570\,A                     & 2.4  & K5III   & 2.10  & 2.04$\pm$0.02   \\
$\chi$ Sco   &  GJ\,570\,A, GJ\,663\,A  & 20.6, 20.7 &  K3III  & 2.09  & 2.04$\pm$0.02   \\
\hline

               \end{tabular}
         \label{tab-3}
   \end{table*}

\begin{table*}
    \caption[]{List of calibrator stars used during AMBER runs. Angular diameters come from \citet{Merand2004}}
      \begin{tabular}{llclcc}
\hline
Calibrator & Target & Ang. dist. & Spectral & $m_{K}$ & $\theta_{UD}$ K band  \\
                    &              & degrees & Type     &     & [$mas$]                           \\
\hline \noalign{\smallskip}	
HD\,25700  &   GJ\,166\,A  & 9.3 & K3III   & 3.16 & 1.04$\pm$0.01  \\
HD\,27508 &    GJ\,166\,A  & 9.9 & K5III   & 3.60 & 0.98$\pm$0.01  \\
HD\,127897 &  GJ\,551    & 10.2 &  K4III  & 3.82 &  0.91$\pm$0.01  \\
HD\,128713 &  GJ\,551    & 6.4 & K0.5II & 3.49 & 0.86$\pm$0.01   \\
HD\,130227 &  GJ\,551    & 6.5 & K1III   & 3.59 & 0.92$\pm$0.01   \\
HD\,136289 &  GJ\,551    & 10.1 &  K3III  & 3.54 & 0.94$\pm$0.01   \\
HD\,204609 &  GJ\,879    & 19.9 & K7III   & 3.20 & 1.14$\pm$0.02   \\
HD\,205096 &  GJ\,879    & 19.8 &  K1III  & 3.82 & 0.82$\pm$0.01   \\
HD\,215627 &  GJ\,879, GJ\,887 & 10.3, 6.9 & K2III & 3.91 & 0.83$\pm$0.01   \\
HD\,221370 &  GJ\,887    & 7.8 &  K2III  & 3.61 & 0.90$\pm$0.01   \\
HD\,223428 &  GJ\,879    & 12.4 & K1III   & 3.21 & 1.07$\pm$0.01   \\
\hline
         \end{tabular}
         \label{tab-2}
   \end{table*}


 \begin{table*}
      \caption{Stellar properties. Masses for M-dwarfs are derived from \citet{Delfosse2000} while empirical relation from \citet{Xia2008} is used for K-dwarfs. Metallicity values provided are all directly determined by spectroscopy.}
               \label{tab-6}
         \begin{tabular}{llccccccccclc}
           \hline
            Star &Spect. &$M_{K}$& Teff   & Ref.&  Mass  & [Fe/H] & Ref.& Derived $T_{eff}$ & $v$sin$i$ & log $R'_{HK}$& Instr. &  Radius\\  
                    &  Type    &               &    K     & & [$M_{\odot}$]    &    &               &  K             & km/s     &             &    & [$R_{\odot}$] \\  
\hline \noalign{\smallskip}
            GJ\,663\,A & K0V    &  -       &      -      &     &           -                                               &  -0.20 &(3)         & 4843$\pm$134    &   -         &    -      & VINCI      & 0.817$\pm$0.016 \\  
            GJ\,166\,A & K0.5V &  3.90$\pm$0.02&  5201 &(3) &   0.877$\pm$0.044      & -0.25 &(3)         & 5269$\pm$35      &   0.78    &  -4.87  & AMBER   & 0.770$\pm$0.021\\  
            GJ\,570\,A & K4V   &  4.20$\pm$0.03&  4758 &(3)  &   0.802$\pm$0.040      &   0.06 &(3)         & 4597$\pm$101    & 1.50     & -4.48 &  VINCI     &0.739$\pm$0.019  \\  
            GJ\,845   & K5V     &  4.38$\pm$0.03&  4630 &(3)   &   0.762$\pm$0.038      & -0.06 &(3)          & 4568$\pm$59      &  1.46    & -4.56 &  VINCI     &0.732$\pm$0.006  \\ 
            GJ\,879   & K5Vp   &  4.54$\pm$0.08&  4574 &(3)   &   0.725$\pm$0.036      &  0.02 &(3)          & 4711 $\pm$134   &  2.93    & -4.27 &  AMBER & 0.629$\pm$0.051   \\ 
	   GJ\,887   & M0.5V &  5.78$\pm$0.03& 3626 &(1)    &  0.503$\pm$0.025 &       -0.22  &(4)        & 3797 $\pm$45      &   -         &  -        &  AMBER  & 0.459$\pm$0.011 \\                     
            GJ\,551   & M5.5V &  8.80$\pm$0.04& 3042 &(1)    &  0.123$\pm$0.006  &  0.19        &(2)        & 3098$\pm$56        &  -        &   -         & AMBER & 0.141$\pm$0.007\\ 
           \hline
         \end{tabular}
         \begin{itemize}{}{}
         \item References:
\item For $T_{\rm eff}$ and [Fe/H]:
(1) \citet{Segransan2003}; (2) \citet{Edvardsson1993}, (3) CORALIE and (4) \citet{Woolf2005}.
\item For $v$sin$i$ and log $R'_{HK}$:
CORALIE
\end{itemize}
   \end{table*}


 \begin{table*}
      \caption[]{Derived uniform disk and limb-darkened diameters, 
                 and stellar radii. Parallaxes are from Hipparcos 
                 \citep{Hipparcos}}. 
               \label{tab-5}
         \begin{tabular}{llcccccc}
           \hline
             Target & Instrument    &  parallax     &$m_{K}$ & Limb darkening coeff. & $\theta_{UD}$& $\theta_{LD}$ \\                  
		       &                        &  [$mas$]      &                &            K band                &       [$mas$]     &     [$mas$]             \\
\hline \noalign{\smallskip}	
            GJ\,663\,A& VINCI       &168.54$\pm$0.54 &                 -                &  [0.79, -0.56, 0.43, -0.15]  &1.253$\pm$0.025 & 1.282$\pm$0.026 \\ 
            GJ\,166\,A & AMBER       &200.62$\pm$0.23 & 2.39$\pm$0.02 (a)& [0.81, -0.53, 0.39, -0.13]  &1.405$\pm$0.038 & 1.437$\pm$0.039 \\
            GJ\,570\,A & VINCI      &171.22$\pm$0.94 & 3.15$\pm$0.02 (a) &[0.86, -0.52, 0.37, -0.12]  &1.147$\pm$0.029& 1.177$\pm$0.030 \\  
            GJ\,845  & VINCI        &276.06$\pm$0.28 & 2.18$\pm$0.02 (b)& [0.86, -0.53, 0.38, -0.13]  &1.834$\pm$0.016& 1.881$\pm$0.017 \\ 
            GJ\,879  & AMBER    &131.42$\pm$0.62 & 3.95$\pm$0.08 (a)& [0.86, -0.53, 0.38, -0.13] &0.750$\pm$0.066 & 0.769$\pm$0.067 \\ 
            GJ\,887  & AMBER      &305.26$\pm$0.70 & 3.36$\pm$0.02 (a)& [1.61, -2.35, 2.00, -0.68]  &1.284$\pm$0.031 & 1.304$\pm$0.032 \\
            GJ\,551  & AMBER      &771.64$\pm$2.60 & 4.38$\pm$0.03 (c)&  [1.94, -2.80, 2.39, -0.81]  &0.990$\pm$0.050 & 1.011$\pm$0.052 \\ 
           \hline
         \end{tabular}
                  \begin{itemize}{}{}
         \item References:
         (a) \citet{Morel1978}; (b) \citet{Mould1976}; (c) 2MASS \citep{Cutri2003} and (d) \citet{Segransan2003}.
\end{itemize}
   \end{table*}


\bibliographystyle{aa}
\bibliography{vlti-mass-radii}

\begin{thebibliography}{42}
\expandafter\ifx\csname natexlab\endcsname\relax\def\natexlab#1{#1}\fi

\bibitem[{Baraffe {et~al.}(1998)Baraffe, Chabrier, Allard, \&
  Hauschildt}]{Baraffe1998}
Baraffe, I., Chabrier, G., Allard, F., \& Hauschildt, P.~H. 1998, Astronomy and
  Astrophysics, 337, 403

\bibitem[{Berger {et~al.}(2006)Berger, Gies, McAlister, ten Brummelaar, Henry,
  Sturmann, Sturmann, Turner, Ridgway, Aufdenberg, \& M{\'e}rand}]{Berger2006}
Berger, D.~H., Gies, D.~R., McAlister, H.~A., {et~al.} 2006, The Astrophysical
  Journal, 644, 475, (c) 2006: The American Astronomical Society

\bibitem[{{B{\"o}hm-Vitense}(1958)}]{Bohm1958}
{B{\"o}hm-Vitense}, E. 1958, Zeitschrift fur Astrophysik, 46, 108

\bibitem[{{Bonfils} {et~al.}(2005){Bonfils}, {Delfosse}, {Udry}, {Santos},
  {Forveille}, \& {S{\'e}gransan}}]{Bonfils2005}
{Bonfils}, X., {Delfosse}, X., {Udry}, S., {et~al.} 2005, \aap, 442, 635

\bibitem[{{Bord{\'e}} {et~al.}(2002){Bord{\'e}}, {Coud{\'e} du Foresto},
  {Chagnon}, \& {Perrin}}]{Borde2002}
{Bord{\'e}}, P., {Coud{\'e} du Foresto}, V., {Chagnon}, G., \& {Perrin}, G.
  2002, \aap, 393, 183

\bibitem[{{Boyajian} {et~al.}(2008){Boyajian}, {McAlister}, {Baines}, {Gies},
  {Henry}, {Jao}, {O'Brien}, {Raghavan}, {Touhami}, {ten Brummelaar},
  {Farrington}, {Goldfinger}, {Sturmann}, {Sturmann}, {Turner}, \&
  {Ridgway}}]{Boyajian2008}
{Boyajian}, T.~S., {McAlister}, H.~A., {Baines}, E.~K., {et~al.} 2008, \apj,
  683, 424

\bibitem[{Chabrier {et~al.}(2007)Chabrier, Gallardo, \& Baraffe}]{Chabrier2007}
Chabrier, G., Gallardo, J., \& Baraffe, I. 2007, A{\&}A, 472, L17

\bibitem[{Claret(2000)}]{Claret2000}
Claret, A. 2000, A{\&}A, 363, 1081

\bibitem[{{Coude Du Foresto} {et~al.}(1998){Coude Du Foresto}, {Perrin},
  {Ruilier}, {Mennesson}, {Traub}, \& {Lacasse}}]{1998SPIE.3350..856C}
{Coude Du Foresto}, V., {Perrin}, G., {Ruilier}, C., {et~al.} 1998, in
  Presented at the Society of Photo-Optical Instrumentation Engineers (SPIE)
  Conference, Vol. 3350, Society of Photo-Optical Instrumentation Engineers
  (SPIE) Conference Series, ed. R.~D. {Reasenberg}, 856--863

\bibitem[{{Cutri} {et~al.}(2003){Cutri}, {Skrutskie}, {van Dyk}, {Beichman},
  {Carpenter}, {Chester}, {Cambresy}, {Evans}, {Fowler}, {Gizis}, {Howard},
  {Huchra}, {Jarrett}, {Kopan}, {Kirkpatrick}, {Light}, {Marsh}, {McCallon},
  {Schneider}, {Stiening}, {Sykes}, {Weinberg}, {Wheaton}, {Wheelock}, \&
  {Zacarias}}]{Cutri2003}
{Cutri}, R.~M., {Skrutskie}, M.~F., {van Dyk}, S., {et~al.} 2003, {2MASS All
  Sky Catalog of point sources.} (The IRSA 2MASS All-Sky Point Source Catalog,
  NASA/IPAC Infrared Science
  Archive.~http://irsa.ipac.caltech.edu/applications/Gator/)

\bibitem[{Davis {et~al.}(2000)Davis, Tango, \& Booth}]{Davis2000}
Davis, J., Tango, W.~J., \& Booth, A.~J. 2000, Monthly Notices RAS, 318, 387,
  (c) 2000 The Royal Astronomical Society

\bibitem[{Delfosse {et~al.}(2000)Delfosse, Forveille, S{\'e}gransan, Beuzit,
  Udry, Perrier, \& Mayor}]{Delfosse2000}
Delfosse, X., Forveille, T., S{\'e}gransan, D., {et~al.} 2000, A{\&}A, 364, 217

\bibitem[{{di Folco} {et~al.}(2007){di Folco}, {Absil}, {Augereau},
  {M{\'e}rand}, {Coud{\'e} Du Foresto}, {Th{\'e}venin}, {Defr{\`e}re},
  {Kervella}, {Ten Brummelaar}, {McAlister}, {Ridgway}, {Sturmann}, {Sturmann},
  \& {Turner}}]{diFolco2007}
{di Folco}, E., {Absil}, O., {Augereau}, J.-C., {et~al.} 2007, \aap, 475, 243

\bibitem[{{Ducati}(2002)}]{Ducati2002}
{Ducati}, J.~R. 2002, VizieR Online Data Catalog, 2237, 0

\bibitem[{{Edvardsson} {et~al.}(1993){Edvardsson}, {Andersen}, {Gustafsson},
  {Lambert}, {Nissen}, \& {Tomkin}}]{Edvardsson1993}
{Edvardsson}, B., {Andersen}, J., {Gustafsson}, B., {et~al.} 1993, \aap, 275,
  101

\bibitem[{{Glass}(1974)}]{Glass74}
{Glass}, I.~S. 1974, Monthly Notes of the Astronomical Society of South Africa,
  33, 53

\bibitem[{Henry \& McCarthy(1993)}]{Henry1993}
Henry, T.~J. \& McCarthy, D.~W. 1993, Astronomical Journal (ISSN 0004-6256),
  106, 773

\bibitem[{{Kervella} {et~al.}(2000){Kervella}, {Coude du Foresto},
  {Glindemann}, \& {Hofmann}}]{2000SPIE.4006...31K}
{Kervella}, P., {Coude du Foresto}, V., {Glindemann}, A., \& {Hofmann}, R.
  2000, in Presented at the Society of Photo-Optical Instrumentation Engineers
  (SPIE) Conference, Vol. 4006, Society of Photo-Optical Instrumentation
  Engineers (SPIE) Conference Series, ed. P.~{L{\'e}na} \& A.~{Quirrenbach},
  31--42

\bibitem[{{Kervella} {et~al.}(2008){Kervella}, {M{\'e}rand}, {Pichon},
  {Th{\'e}venin}, {Heiter}, {Bigot}, {Ten Brummelaar}, {McAlister}, {Ridgway},
  {Turner}, {Sturmann}, {Sturmann}, {Goldfinger}, \&
  {Farrington}}]{Kervella2008}
{Kervella}, P., {M{\'e}rand}, A., {Pichon}, B., {et~al.} 2008, \aap, 488, 667

\bibitem[{Kervella {et~al.}(2004{\natexlab{a}})Kervella, S{\'e}gransan, \&
  du~Foresto}]{KervellaSegransan2004}
Kervella, P., S{\'e}gransan, D., \& du~Foresto, V.~C. 2004{\natexlab{a}},
  A{\&}A, 425, 1161

\bibitem[{Kervella {et~al.}(2004{\natexlab{b}})Kervella, Th{\'e}venin, DiFolco,
  \& S{\'e}gransan}]{Kervella2004}
Kervella, P., Th{\'e}venin, F., DiFolco, E., \& S{\'e}gransan, D.
  2004{\natexlab{b}}, A{\&}A, 426, 297

\bibitem[{Lane {et~al.}(2001)Lane, Boden, \& Kulkarni}]{Lane2001}
Lane, B.~F., Boden, A.~F., \& Kulkarni, S.~R. 2001, The Astrophysical Journal,
  551, L81, (c) 2001: The American Astronomical Society

\bibitem[{{Le Bouquin} {et~al.}(2008){Le Bouquin}, {Abuter}, {Bauvir},
  {Bonnet}, {Haguenauer}, {di Lieto}, {Menardi}, {Morel}, {Rantakyr{\"o}},
  {Schoeller}, {Wallander}, \& {Wehner}}]{LeBouquin2008}
{Le Bouquin}, J.-B., {Abuter}, R., {Bauvir}, B., {et~al.} 2008, in Society of
  Photo-Optical Instrumentation Engineers (SPIE) Conference Series, Vol. 7013,
  Society of Photo-Optical Instrumentation Engineers (SPIE) Conference Series

\bibitem[{L{\'o}pez-Morales(2007)}]{LopezMorales2007}
L{\'o}pez-Morales, M. 2007, The Astrophysical Journal, 660, 732

\bibitem[{{Maceroni} \& {Montalb{\'a}n}(2004)}]{Maceroni2004}
{Maceroni}, C. \& {Montalb{\'a}n}, J. 2004, \aap, 426, 577

\bibitem[{{Malbet} {et~al.}(2008){Malbet}, {Duvert}, {Chelli}, \&
  {Kern}}]{ATF2008}
{Malbet}, F., {Duvert}, G., {Chelli}, A., \& {Kern}, P. 2008, ArXiv e-prints

\bibitem[{Merand {et~al.}(2004)Merand, Borde, \& Foresto}]{Merand2004}
Merand, A., Borde, P., \& Foresto, V. C.~D. 2004, New Frontiers in Stellar
  Interferometry, 5491, 1185

\bibitem[{{Morel} \& {Magnenat}(1978)}]{Morel1978}
{Morel}, M. \& {Magnenat}, P. 1978, \aaps, 34, 477

\bibitem[{{Mould} \& {Hyland}(1976)}]{Mould1976}
{Mould}, J.~R. \& {Hyland}, A.~R. 1976, \apj, 208, 399

\bibitem[{Perrin(2003)}]{Perrin2003}
Perrin, G. 2003, A{\&}A, 400, 1173

\bibitem[{Petrov {et~al.}(2007)Petrov, Malbet, Weigelt, Antonelli, Beckmann,
  Bresson, Chelli, Dugu{\'e}, Duvert, Gennari, Gl{\"u}ck, Kern, Lagarde,
  Coarer, Lisi, Millour, Perraut, Puget, Rantakyr{\"o}, Robbe-Dubois, Roussel,
  Salinari, Tatulli, Zins, Accardo, Acke, Agabi, Altariba, Arezki, Aristidi,
  Baffa, Behrend, Bl{\"o}cker, Bonhomme, Busoni, Cassaing, Clausse, Colin,
  Connot, Delboulb{\'e}, Souza, Driebe, Feautrier, Ferruzzi, Forveille, Fossat,
  Foy, Fraix-Burnet, Gallardo, Giani, Gil, Glentzlin, Heiden, Heininger,
  Utrera, Hofmann, Kamm, Kiekebusch, Kraus, Contel, Contel, Lesourd, Lopez,
  Lopez, Magnard, Marconi, Mars, Martinot-Lagarde, Mathias, M{\`e}ge, Monin,
  Mouillet, Mourard, Nussbaum, Ohnaka, Pacheco, Perrier, Rabbia, Rebattu,
  Reynaud, Richichi, Robini, Sacchettini, Schertl, Sch{\"o}ller, Solscheid,
  Spang, Stee, Stefanini, Tallon, Tallon-Bosc, Tasso, Testi, Vakili, L{\"u}he,
  Valtier, Vannier, \& Ventura}]{Petrov2007}
Petrov, R.~G., Malbet, F., Weigelt, G., {et~al.} 2007, A{\&}A, 464, 1

\bibitem[{{Ribas}(2003)}]{Ribas2003}
{Ribas}, I. 2003, \aap, 398, 239

\bibitem[{{Ribas} {et~al.}(2008){Ribas}, {Morales}, {Jordi}, {Baraffe},
  {Chabrier}, \& {Gallardo}}]{Ribas2008}
{Ribas}, I., {Morales}, J.~C., {Jordi}, C., {et~al.} 2008, Memorie della
  Societa Astronomica Italiana, 79, 562

\bibitem[{{S{\'e}gransan} {et~al.}(2000){S{\'e}gransan}, {Delfosse},
  {Forveille}, {Beuzit}, {Udry}, {Perrier}, \& {Mayor}}]{Segransan2000}
{S{\'e}gransan}, D., {Delfosse}, X., {Forveille}, T., {et~al.} 2000, \aap, 364,
  665

\bibitem[{{S{\'e}gransan} {et~al.}(1999){S{\'e}gransan}, {Forveille},
  {Millan-Gabet}, \& {Traub}}]{Segransan99}
{S{\'e}gransan}, D., {Forveille}, T., {Millan-Gabet}, C.~P.~R., \& {Traub},
  W.~A. 1999, in Astronomical Society of the Pacific Conference Series, Vol.
  194, Working on the Fringe: Optical and IR Interferometry from Ground and
  Space, ed. S.~{Unwin} \& R.~{Stachnik}, 290--+

\bibitem[{S{\'e}gransan {et~al.}(2003)S{\'e}gransan, Kervella, Forveille, \&
  Queloz}]{Segransan2003}
S{\'e}gransan, D., Kervella, P., Forveille, T., \& Queloz, D. 2003, A{\&}A,
  397, L5

\bibitem[{{Soubiran} {et~al.}(2008){Soubiran}, {Bienaym{\'e}}, {Mishenina}, \&
  {Kovtyukh}}]{Soubiran}
{Soubiran}, C., {Bienaym{\'e}}, O., {Mishenina}, T.~V., \& {Kovtyukh}, V.~V.
  2008, \aap, 480, 91

\bibitem[{Tatulli {et~al.}(2007)Tatulli, Millour, Chelli, Duvert, Acke, Utrera,
  Hofmann, Kraus, Malbet, M{\`e}ge, Petrov, Vannier, Zins, Antonelli, Beckmann,
  Bresson, Dugu{\'e}, Gennari, Gl{\"u}ck, Kern, Lagarde, Coarer, Lisi, Perraut,
  Puget, Rantakyr{\"o}, Robbe-Dubois, Roussel, Weigelt, Accardo, Agabi,
  Altariba, Arezki, Aristidi, Baffa, Behrend, Bl{\"o}cker, Bonhomme, Busoni,
  Cassaing, Clausse, Colin, Connot, Delboulb{\'e}, Souza, Driebe, Feautrier,
  Ferruzzi, Forveille, Fossat, Foy, Fraix-Burnet, Gallardo, Giani, Gil,
  Glentzlin, Heiden, Heininger, Kamm, Kiekebusch, Contel, Contel, Lesourd,
  Lopez, Lopez, Magnard, Marconi, Mars, Martinot-Lagarde, Mathias, Monin,
  Mouillet, Mourard, Nussbaum, Ohnaka, Pacheco, Perrier, Rabbia, Rebattu,
  Reynaud, Richichi, Robini, Sacchettini, Schertl, Sch{\"o}ller, Solscheid,
  Spang, Stee, Stefanini, Tallon, Tallon-Bosc, Tasso, Testi, Vakili, L{\"u}he,
  Valtier, \& Ventura}]{Tatulli2007}
Tatulli, E., Millour, F., Chelli, A., {et~al.} 2007, A{\&}A, 464, 29

\bibitem[{Torres \& Ribas(2002)}]{Torres2002}
Torres, G. \& Ribas, I. 2002, The Astrophysical Journal, 567, 1140

\bibitem[{{van Leeuwen}(2007)}]{Hipparcos}
{van Leeuwen}, F., ed. 2007, Astrophysics and Space Science Library, Vol. 250,
  {Hipparcos, the New Reduction of the Raw Data}

\bibitem[{{Woolf} \& {Wallerstein}(2005)}]{Woolf2005}
{Woolf}, V.~M. \& {Wallerstein}, G. 2005, \mnras, 356, 963

\bibitem[{Xia {et~al.}(2008)Xia, Ren, \& Fu}]{Xia2008}
Xia, F., Ren, S., \& Fu, Y. 2008, Astrophys Space Sci, 314, 51, (c) 2008:
  Springer Science+Business Media B.V.

\end{thebibliography}

\appendix
\section{Online Material - Figures}

      \begin{figure*}
  \centering 
    \includegraphics[width=8cm]{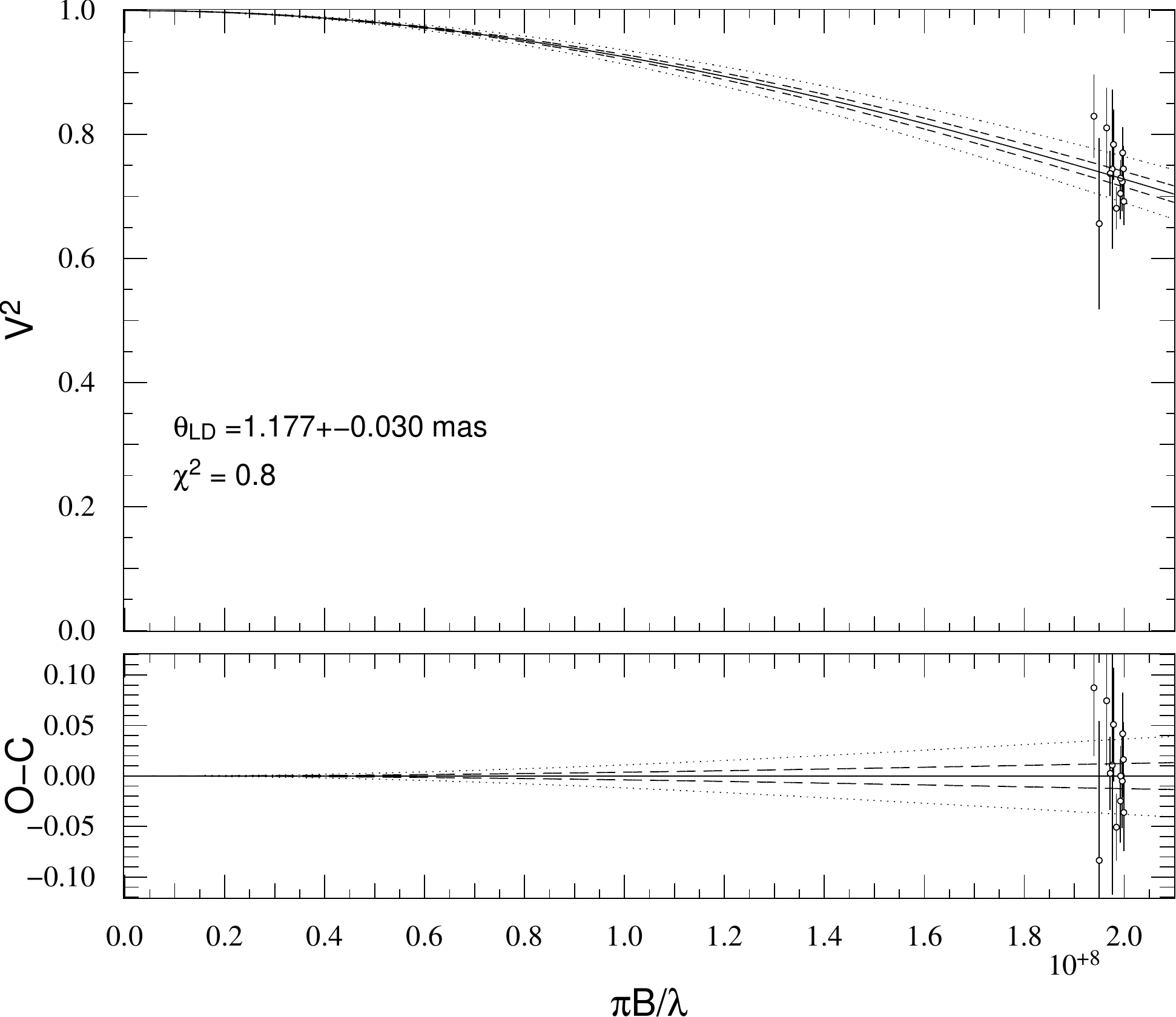}
   \caption{Calibrated squared visibilities from VINCI and
            best-fit LD disk model (solid) for GJ570A vs. spatial frequencies. 1-$\sigma$ (dash) and 3-$\sigma$ (dot) uncertainties are also indicated.
            }
   \label{Fig-2}
    \end{figure*}

          \begin{figure*}
  \centering 
    \includegraphics[width=8cm]{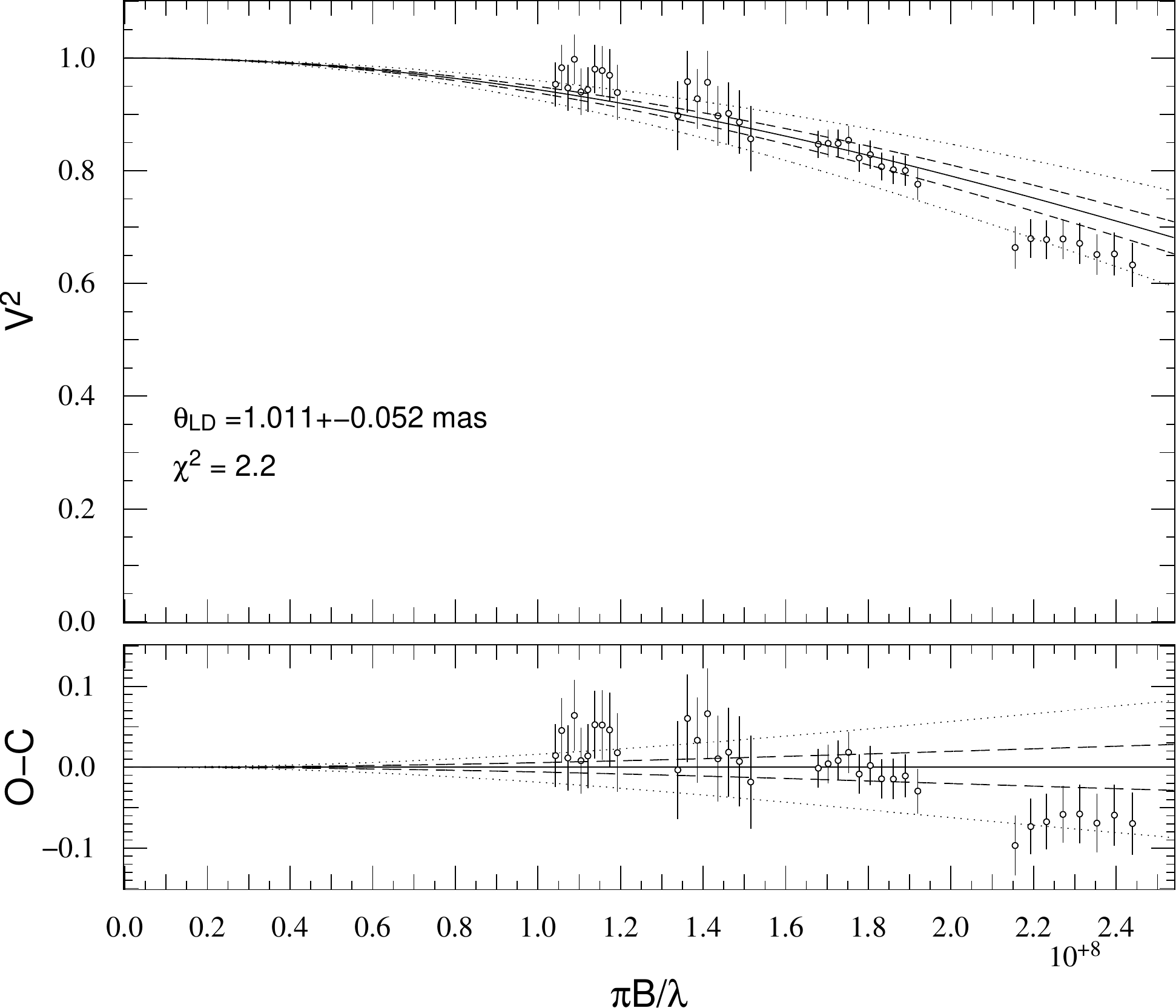}
   \caption{Calibrated squared visibilities from AMBER (low-resolution mode) and
            best-fit LD disk model (solid) for GJ551 vs. spatial frequencies.  1-$\sigma$ (dash) and 3-$\sigma$ (dot) uncertainties are also indicated. Error bars amplitudes include both correlated and non-correlated errors.
            }
   \label{Fig-7}
    \end{figure*}
    
          \begin{figure*}
  \centering 
    \includegraphics[width=8cm]{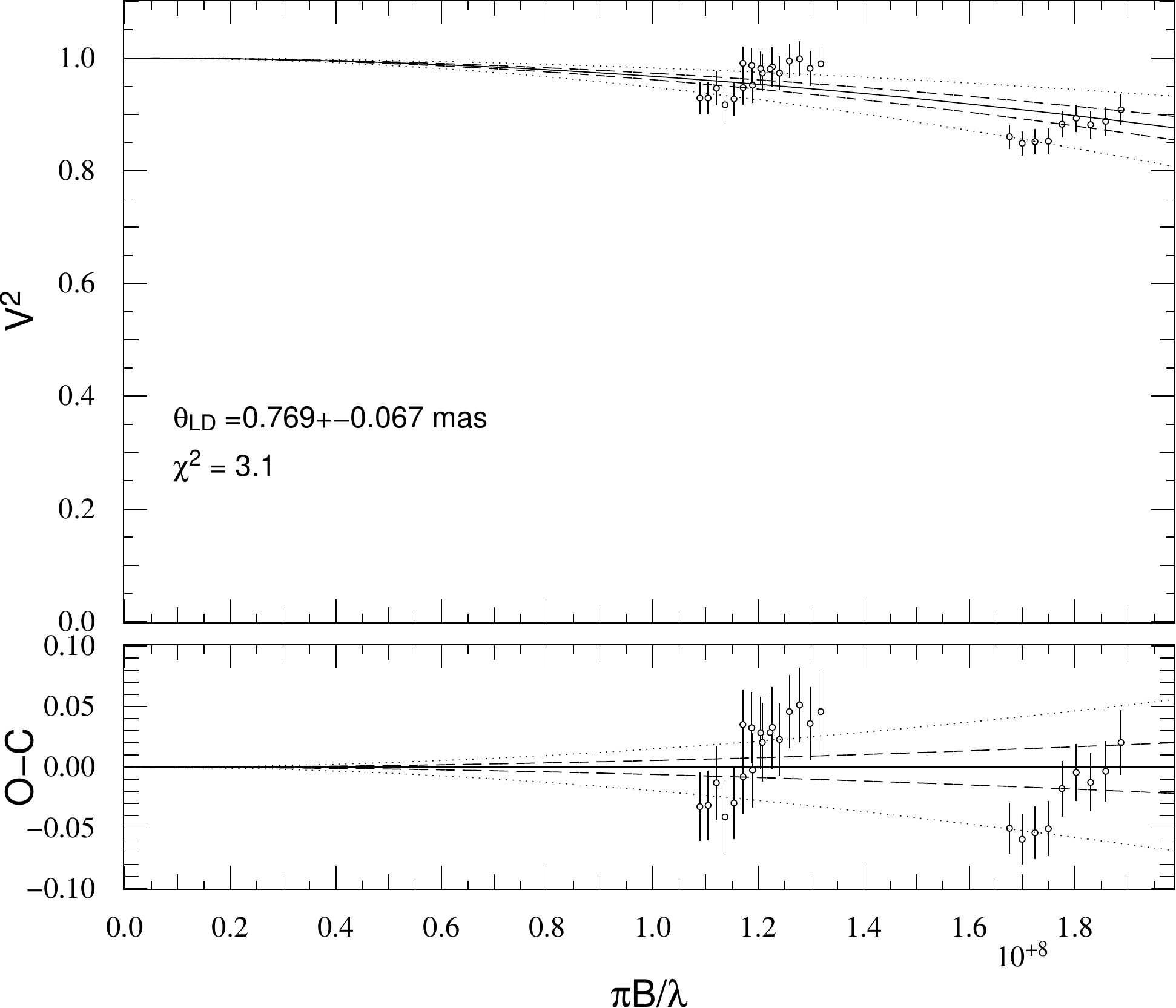}
   \caption{Calibrated squared visibilities from AMBER (low-resolution mode) and
            best-fit LD disk model (solid) for GJ879 vs. spatial frequencies.  1-$\sigma$ (dash) and 3-$\sigma$ (dot) uncertainties are also indicated. Error bars amplitudes include both correlated and non-correlated errors.
            }
   \label{Fig-8}
    \end{figure*}

\section{Online Material - Tables}

   
\begin{table*}
\caption{Online material - Angular diameters and linear radii of single low and very-low mass stars from the literature. They are respectively expressed in mas and sun radii.}
\label{interf_table}
\begin{tabular}{llrclcccc}
\hline
Star & Spect. & [Fe/H] & Ref.$_{[Fe/H]}$ & Instr. &  $\theta_{\rm UD}$ & $\theta_{\rm LD}$ & $R_{lin}$ & Ref.$_{interf.}$\\
\hline \noalign{\smallskip}
$\sigma$\,Dra&K0V &  -0.21    &(a)      &CHARA& $1.224\pm{0.011}$ & $1.254\pm{0.012}$& 0.778$\pm$0.008 &(4)\\ 
HR\,511          &K0V &   0.02  & (a)      &CHARA& $0.747\pm{0.021}$ & $0.763\pm{0.021}$& 0.819$\pm$0.024 &(4)\\ 
$\epsilon$\,Eri&K2V &  -0.06  &(a) &VINCI   & $2.093\pm{0.029}$ & $2.148\pm{0.029}$& 0.735$\pm$0.005 &(1)\\ 
GJ\,105\,A      &K3V &    -0.07  &(a)  &PTI        &  $0.914\pm{0.07}$   & $0.936\pm{0.07}$  &0.708$\pm$0.050 &(2) \\
61\,Cyg\,A       &K5V&    -0.27  &(a)   &CHARA&                                & $1.775\pm{0.013}$& 0.665$\pm$0.005 &(5)\\
GJ\,380           &K7V&    -0.03  &(a)         &PTI        &  $1.268\pm{0.04}$ & $1.155\pm{0.04}$&0.605$\pm$0.020&(2) \\
61\,Cyg\,B      &K7V&     -0.39  &(a)            &CHARA&                                & $1.581\pm{0.022}$& 0.595$\pm$0.008 &(5)\\
GJ\,191          &M1V  &    -0.50  &(b)         &VINCI    &  $0.681\pm{0.06}$ & $0.692\pm{0.06}$&0.291$\pm$0.025&(3) \\
GJ\,887           &M0.5V&    -0.22  &(c)      &VINCI       & $1.366\pm{0.04}$ & $1.388\pm{0.04}$&0.491$\pm$0.014&(3) \\
GJ\,205           &M1.5V &     0.21  &(c)           &VINCI    & $1.124\pm{0.11}$ & $1.149\pm{0.11}$&0.702$\pm$0.063 &(3)\\
GJ\,15\,A         &M2V  &    -0.45  &(d)           &PTI       & $0.976\pm{0.016}$ & $0.988\pm{0.016}$&0.379$\pm$0.006&(6) \\
GJ\,411           &M1.5V &    -0.31  &(a)          &PTI        & $1.413\pm{0.03}$ & $1.436\pm{0.03}$&0.393$\pm$0.008&(2) \\
GJ\,699           &M4Ve &    -0.15  &(a)           &PTI       & $0.987\pm{0.04}$ & $1.004\pm{0.04}$&0.196$\pm$0.008 &(2)\\
GJ\,551          &M5.5V  &     0.19   &(f)          &VINCI     & $1.023\pm{0.08}$ & $1.044\pm{0.08}$&0.145$\pm$0.011&(3) \\
\hline
\end{tabular}
\begin{itemize}{}{}
\item Ref. for [Fe/H]:
(a) \citet{Soubiran}; 
(b) \citet{Mould1976}; 
(c) \citet{Woolf2005}; 
(d) \citet{Bonfils2005}; 
(e) \citet{Edvardsson1993}.

\item Ref. for interferometric measurements:
(1) \citet{diFolco2007}; 
(2) \citet{Lane2001}; 
(3) \citet{Segransan2003}; 
(4) \citet{Boyajian2008}; 
(5) \citet{Kervella2008}; 
(6) \citet{Berger2006}.
\end{itemize}
\end{table*}

   
\begin{table*}
\caption{
Apparent magnitudes of stars included in this study.
The uncertainty adopted for each apparent magnitude value is given in superscript.
}
\label{phot_table}
\begin{tabular}{lrrrrrrrrr}
\hline
Star & $m_U$$^{(a)}$ & $m_B$$^{(b)}$ & $m_V$$^{(b)}$ & $m_R$$^{(c)}$ & $m_I$$^{(c)}$ & $m_J$$^{(d)}$ & $m_H$$^{(d)}$ & $m_K$$^{(d)}$ & $m_L$$^{(d)}$ \\
\hline
\noalign{\smallskip}
\object{GJ\,166\,A} & $5.69^{0.02}$ & $5.25^{0.02}$ & $4.43^{0.02}$ & $3.72^{0.01}$ & $3.27^{0.01}$ & $2.91^{0.03}$ & $2.46^{0.01}$ & $2.39^{0.02}$ & $2.30^{0.02}$\\
\object{GJ\,663\,A} & $5.67^{0.02}$ & $5.18^{0.02}$ & $4.32^{0.02}$ & $3.62^{0.02}$ & $3.18^{0.02}$ &  &  &  & \\ 
\object{GJ\,570\,A} & $7.88^{0.02}$ & $6.82^{0.02}$ & $5.71^{0.02}$ & $4.72^{0.02}$ & $4.18^{0.02}$ & $3.82^{0.02}$ & $3.27^{0.02}$ & $3.15^{0.02}$ & $3.11^{0.02}$\\
\object{GJ\,845} & $6.74^{0.02}$ & $5.75^{0.02}$ & $4.69^{0.02}$ & $3.81^{0.02}$ & $3.25^{0.02}$ & $2.83^{0.02}$ & $2.30^{0.02}$ & $2.18^{0.02}$ & $2.12^{0.02}$\\
\object{GJ\,879} & $8.61^{0.02}$ & $7.59^{0.02}$ & $6.49^{0.02}$ & $5.54^{0.02}$ & $4.95^{0.02}$ & $4.53^{0.04}$ & $4.00^{0.07}$ & $3.95^{0.08}$ & \\ 
\object{GJ\,887} & $9.99^{0.02}$ & $8.83^{0.02}$ & $7.35^{0.02}$ &  &  & $4.20^{0.02}$ & $3.60^{0.02}$ & $3.36^{0.02}$ & $3.20^{0.02}$\\
\object{GJ\,551} & $14.56^{0.02}$ & $13.02^{0.02}$ & $11.05^{0.02}$ & $8.68^{0.02}$ & $6.42^{0.02}$ & $5.33^{0.02}$ & $4.73^{0.02}$ & $4.36^{0.03}$ & $4.04^{0.02}$\\
\hline
\end{tabular}
\begin{itemize}{}{}
\item References:
\item[$(a)$] \citet{Morel1978}.
\item[$(b)$] \citet{Morel1978}, \citet{Hipparcos}. 
\item[$(c)$] \citet{Morel1978}, \citet{Ducati2002}.
\item[$(d)$] \citet{Morel1978},\citet{Glass74}, \citet{Mould1976}, \citet{Ducati2002},\citet{Cutri2003}.

\end{itemize}
\end{table*}

\end{document}